%% file: CompositionalBioChemistryModel.tex
\documentclass[preprint,12pt]{article}
\usepackage[table]{xcolor}
\usepackage[margin=1in]{geometry}

\usepackage{tikz}
\usetikzlibrary{calc, matrix}
\usetikzlibrary{arrows}  
\usetikzlibrary{positioning, shapes}
\usetikzlibrary{backgrounds}
\usetikzlibrary{svg.path}
\usetikzlibrary{decorations.markings, decorations.pathreplacing}
\usetikzlibrary{graphs}
\usetikzlibrary{bending}

\usepackage{chemformula}

\usepackage{graphicx} 
\usepackage{amsmath,amsfonts,amssymb,amsthm}
\usepackage[pgf]{adjustbox}
\usepackage{caption}
\usepackage{subcaption}
\usepackage{hyperref}
\usepackage{booktabs}
\hypersetup{colorlinks=false}
\usepackage{comment}

\usepackage[english]{babel}
\usepackage[T1]{fontenc}
\usepackage[utf8]{inputenc}
\usepackage{babel}
\usepackage[font=small,labelfont=bf]{caption}
\usepackage[numbers]{natbib}
\setcitestyle{authoryear,open={(},close={)}} 
\usepackage{tikz}
\usepackage{verbatim}
\usepackage{siunitx}
\usepackage{comment}
\usepackage{soul}

\DeclareSIUnit\darcy{D}
\DeclareSIUnit\poise{P}
\DeclareSIUnit\atm{atm}
\DeclareSIUnit\smeter{sm}
\DeclareSIUnit\bar{bar}
\DeclareSIUnit\kgw{kgw}
\usepackage[version=4]{mhchem}
\newcommand{\chemset}[1]{i \in \{\ce{#1}\}}
\newcommand{\waterphase}{l}
\newcommand{\gasphase}{g}

\newcommand{\msalt}{m_{salt}}

\newcommand{\bse}{\begin{subequations}}
\newcommand{\ese}{\end{subequations}}
\newcommand{\Phiref}{\Phi_{\textnormal{ref}}}
\newcommand{\Aalpha}{A_{\zeta}}

\newcommand{\Salpha}{S_{\zeta}}
\newcommand{\Xalphai}{X_{\zeta,i}}
\newcommand{\Xalphaj}{X_{\zeta,j}}

\newcommand{\rhot}{\rho_{t}}
\newcommand{\bactN}{\mathcal{N}}

\newcommand{\zi}{z_{i}}

\newcommand{\Tc}{T_{\text{c}}} 
\newcommand{\Pc}{P_{\text{c}}} 
\newcommand{\BIP}{k} 
\newcommand{\acf}{\omega} 
\newcommand{\molmass}{M} 
\newcommand{\molarvol}{V} 
\newcommand{\gasconst}{R} 
\newcommand{\temp}{T} 
\newcommand{\press}{P} 
\newcommand{\comp}{X} 
\newcommand{\eoscoeffa}{a} 
\newcommand{\eoscoeffb}{b} 
\newcommand{\eosalpha}{\alpha} 

\newcommand{\psigrowthmax}{\psi_{\text{max}}^{\text{growth}}}
\newcommand{\psigrowth}{\psi^{\text{growth}}}
\newcommand{\psidecay}{\psi^{\text{decay}}}
\newcommand{\bdecay}{b_{\text{decay}}}
\newcommand{\YHtwo}{Y_{\ce{H2}}}

\newcommand{\lHtwo}{l_{\ce{H2}}}
\newcommand{\lCOtwo}{l_{\ce{CO2}}}

\newcommand{\xHtwol}{X_{l,\ce{H2}}}
\newcommand{\xCOtwol}{X_{l,\ce{CO2}}}

\newcommand{\Sl}{S_{l}}

\newcommand{\palpha}{p_{\zeta}}
\newcommand{\salpha}{S_{\zeta}}

\newcommand{\rhoalpha}{\rho_{\zeta}}

\newcommand{\qi}{q_{i}}
\newcommand{\vK}{\mathbf{K}}

\newcommand{\pclg}{p_{c,\waterphase \gasphase}}

\newcommand{\pg}{p_{\gasphase}}
\newcommand{\po}{p_{\waterphase}}

\newcommand{\ui}{{\mathbf{u}}_{i}}
\newcommand{\jialpha}{{\mathbf{j}}_{\zeta,i}}

\newcommand{\valpha}{{\mathbf{v}}_{\zeta}}

\newcommand{\divec}{\nabla{\cdot}}

\usepackage{authblk}

\usepackage{pdfpages}

\graphicspath{{./},{./figs/}}
\graphicspath{{./},{./fig/}}
\graphicspath{{./},{./fig_bench/}}
\title{Modeling and Simulation of Coupled Biochemical and Two-Phase Compositional Flow in Underground Hydrogen Storage}

\author[1]{Elyes Ahmed\thanks{Elyes.Ahmed@sintef.no}}
\author[2]{Brahim Amaziane\thanks{brahim.amaziane@univ-pau.fr}}
\author[3,4]{Salaheddine Chabab\thanks{salaheddine.chabab@univ-pau.fr}}
\author[2]{Stéphanie Delage Santacreu\thanks{stephanie.delage@univ-pau.fr}}
\author[3]{Guillaume Galliéro\thanks{guillaume.galliero@univ-pau.fr}}
\author[1]{Olav Møyner\thanks{Olav.Moyner@sintef.no}}
\author[1]{Xavier Raynaud\thanks{Xavier.Raynaud@sintef.no}}
\affil[1]{SINTEF Digital, Forskningsveien 1, 0373 Oslo, Norway}
\affil[2]{Universit\'{e} de Pau et des Pays de l'Adour, E2S UPPA, CNRS, LMAP, UMR5142, Pau, France}
\affil[3]{Universit\'{e} de Pau et des Pays de l'Adour, E2S UPPA, CNRS, LFCR, UMR5150, Pau, France}
\affil[4]{Universit\'{e} de Pau et des Pays de l'Adour, E2S UPPA, LaTEP, Pau, France}

\input{comment-utils}

\createcomment{xavier}{softyellow}

\begin{document}

\maketitle

\begin{abstract}
Integrating microbial activity into underground hydrogen storage models is crucial for simulating
long-term reservoir behavior. In this work, we present a coupled framework that incorporates bio-geochemical reactions and compositional flow models within the Matlab Reservoir Simulation
Toolbox (MRST). Microbial growth and decay are modeled using a double Monod formulation,
with populations influenced by hydrogen and carbon dioxide availability. First, a refined Equation
of State (EoS) is employed to accurately capture hydrogen dissolution, thereby improving phase behavior and modeling of microbial activity. The model is then discretized using a cell-centered finite-volume method with implicit Euler time discretization. A fully coupled fully implicit strategy is considered. Our implementation builds upon MRST's compositional module by incorporating the Søreide-Whitson EoS, microbial reaction kinetics, and specific effects such as bio-clogging and molecular diffusion. Through a series of 1D, 2D and 3D simulations, we analyze the effects of microbial-induced bio-geochemical transformations on underground hydrogen storage in porous media.These results highlight that accounting for bio-geochemical effects can substantially impact hydrogen loss, purity, and overall storage performance.

\end{abstract}
\vspace{3mm}



\begin{tabular}{lll}
\hline \multicolumn{2}{l}{ Nomenclature } \\
Symbol & Description & Unit \\
$\bdecay$ & Constant decay rate & \unit{\per\second} \\
$\mathbf{D}_{\zeta,i}$ & Effective molecular diffusion coefficient in phase $\zeta$ & \unit{\square\meter\per\second} \\
$\mathbf{D}_{\bactN,l}$ & Microbial diffusion coefficient in the liquid phase & \unit{\square\meter\per\second} \\
$f_{l,i}$ & Fugacity of component $i$ in the liquid phase & \unit{\pascal} \\
$f_{g,i}$ & Fugacity of component $i$ in the gaseous phase & \unit{\pascal} \\
$\mathbf{g}$  & Gravity acceleration & \unit{\meter\per\square\second} \\
$\gamma_i$ & Stoichiometric coefficient of component $i$ & - \\
$\jialpha$ & Diffusive flux of component $i$ in phase $\zeta$ & \unit{\kilogram\per\square\meter\per\second} \\
$\mathbf{K}$  & Absolute permeability & \unit{\milli\darcy} \\
$k_{r\zeta}$  & Relative permeability of phase $\zeta$ &  - \\
$\lHtwo$ & Half-saturation constant for $\ce{H2}$ & \unit{\mole\per\mole} \\
$\lCOtwo$ & Half-saturation constant for $\ce{CO2}$ & \unit{\mole\per\mole} \\
$\molmass$ & Molar mass & \unit{\kilogram\per\mole} \\
$\molmass_i$ & Molar mass of component $i$ & \unit{\kilogram\per\mole} \\
$\msalt$ & Salt molality & \unit{\mole\per\kgw} \\
$\mu_\zeta$ & Dynamic viscosity of phase $\zeta$ & \unit{\pascal\second} \\
$n$  & Number of microorganisms per volume in water & \unit{\per\cubic\meter} \\
$n_0$  & Initial number of microorganisms per volume in water & \unit{\per\cubic\meter} \\
$\bactN$  & Dimensionless number of microorganisms per volume in water & - \\
$p_l$  & Liquid phase pressure & \unit{\pascal} \\
$p_g$  & Gas phase pressure & \unit{\pascal} \\
$\pclg$  & Capillary pressure & \unit{\pascal} \\
$\Pc$ & Critical pressure & \unit{\pascal} \\
$P$ & total pressure & \unit{\pascal} \\
$\mathbf{\Phi}$ & Porosity & - \\
$\psi^{\text{growth}}$ & Growth rate & \unit{\per\second} \\
$\psi^{\text{decay}}$ & Decay rate & \unit{\per\second} \\
$\rho_l$ & Density of the liquid phase & \unit{\kilogram\per\cubic\meter} \\
$\rho_g$ & Density of the gas phase & \unit{\kilogram\per\cubic\meter} \\
$\Sl$ & Liquid phase saturation & - \\
$S_{lr}$ & Residual saturation liquid phase  & - \\
$S_g$ & Gas phase saturation & - \\
$S_{gr}$ & Residual saturation gaseous phase  & - \\
$\mathbf{v}_\zeta$ & Darcy's law velocity of phase $\zeta$ & \unit{\meter\per\second} \\
$\Xalphai$ & Molar fraction of component $i$ in the phase $\zeta$& - \\
\(\YHtwo\) & Yield coefficient for \(\ce{H2}\) & \unit{\per\mole} \\
t & Time & \unit{\second} \\
T & Temperature & \unit{\kelvin} \\
$\Tc$ & Critical temperature & \unit{\kelvin} \\
$\molarvol$ & Molar volume & \unit{\cubic\meter\per\mole} \\
$\acf$ & Acentric factor & - \\
$z_i$ & Overall mole fraction of component $i$ & - \\
\end{tabular}

\section{Introduction}

Faced with an increase in the energy demand and the necessity to reduce the share of fossil fuels in the global energy supply for environmental reasons, several alternatives and methods are explored and developed to produce energy from renewable sources. Some of them, such as solar or wind energy, result in intermittent energies. Hence, the Power-to-Gas concept (\cite{WULF2018309}) appears to be an interesting solution for energy storage to balance supply and demand in energy. In the short term, hydrogen can be stored using conventional methods, either in above-ground metal reservoirs or underground in salt caverns. As demand for hydrogen increases in the medium to long term, massive storage methods will be required. For this purpose, underground hydrogen storage (UHS) in porous reservoirs looks promising and reliable (\cite{ZIVAR2021}).

Hydrogen can be stored in depleted hydrocarbon reservoirs or aquifers in which physical, chemical, and biological elements can interact and evolve in the liquid-gas porous media and then be adequately delivered to meet the needs. 
Therefore, it is essential to determine the hydrogen thermophysical properties (\cite{chabab2020measurements}), to understand the processes that occur in a liquid-gas phase and a porous multicomponent medium, and appreciate the interactions between microorganisms and chemical elements and also their impact on the rock (\cite{Panfilov2018}). Note that only some microorganisms can adapt to these difficult thermodynamical conditions in the subsurface, which consist essentially of some archaea (methanogenic, homoacetogenic) and bacteria (sulfate-reducing and iron(III)-reducing). Although chemical components exist in both phases, microorganisms are present only in the water phase. Both of these microorganisms consume \ce{H2}. In fact, microbial activity can lead to a reduction in \ce{H2} production and alter rock properties through bio-clogging (\citet{Eddaoui_Panfilov2021}).  Moreover, the formation of biofilms by hydrogen-consuming microorganisms can change the flow properties of the reservoir, while microbial-induced mineral precipitation may result in a reduction in localized permeability. Consequently, accurately estimating the hydrogen available in the liquid phase is essential to evaluate the microbial population and understand its impact on the system.

Several models have been developed to account for the impact of chemical and biological factors and processes~\cite{Panfilov2018, Khoshnevis2023}. Experimental studies have been conducted to derive model parameters for microbial models~\cite{ALI2023108405, THAYSEN2021111481} and to calibrate numerical models~\cite{hogeweg2024development, dopffel2023microbial, zhao2025improved}. In this work, we build on previous work and couple a double Monod model with an overall composition formulation, supplemented by the accurate Sreide-Whitson Equation of State (EoS) model and pore-clogging models. We use the MATLAB Reservoir Simulation Toolbox (MRST)~\cite{lie2019introduction}, focusing on the dynamics of methanogenic archaea. It is important to note that the models and methods implemented here can easily be adapted for other microorganisms. A previous study~\cite{AHMED2024104772} used a black-oil model restricted to water and hydrogen within the MRST framework. Although adequate for capturing key physical processes, it cannot account for microbial activity. In this work, we adopt a fully compositional approach, enabling a more accurate thermodynamic description and the inclusion of micro-organisms effects.

The rest of the paper is organized as follows. In Section \ref{section:biochemistrymodel}, we present an overall compositional model for the two-phase multicomponent transport equations coupled to a methanogenic archae model and estimate its effect on rock properties. In Section \ref{section:thermomodels}, we describe the EoS models that can capture gas solubility and, more particularly the Sreide-Whitson model adapted to MRST. The simulation results are compared with experiments when they are available.  Then, in Section~\ref{section:numericalresults}, we present a series of simulation cases designed to evaluate the model's ability to predict the solubility of \ce{H2} and \ce{CO2}, capture microbial dynamics and their impact on multicomponent liquid-gas phase transport, and assess potential losses of \ce{H2}.

\section{Two-phase multi-component transport with microbial activity model}\label{section:biochemistrymodel}
This section presents a two-phase overall-composition flow model incorporating microbial dynamics, particularly methanogenic archaea, for simulating subsurface interactions in underground hydrogen storage. Unlike the molar fraction, natural variable formulation in \cite{hagemann2017-thesis}, which uses mole conservation, our model is based on mass conservation and employs pressure and overall mole fractions for flow, along with normalized micro-organisms population for microbial dynamics, as primary variables.
\subsection{Methanogenic Archae model}
The methanogenic archae are anaerobic microorganisms present in water and able to grow under difficult thermodynamical conditions. They can live in underground reservoirs or aquifers where they play an essential ecological role by consuming substrates such as \ce{H2} and \ce{CO2} and by producing \ce{CH4}. This methanogenis reaction can occur at low temperatures and high pressures with optimum conditions around $T=308.15\,\unit{\kelvin}$ and $P=90\,\unit{\bar}$ (\cite{Panfilov2018,Eddaoui_Panfilov2021,hagemann2017-thesis}) and can be expressed as:
\begin{align}
    \label{bioreaction}
    4 \ce{H2}+\ce{CO2} \rightarrow \ce{CH4}+2 \ce{H2O}
\end{align}
The microbial dynamics are governed by a growth rate function, $\psigrowth$, and a decay rate function, $\psidecay$, leading to the following conservation equation \cite{Panfilov2018,hagemann2017-thesis,eddaoui2021}:  
\begin{align}\label{eq:bactmodel}
&\frac{\partial (\Phi \rho_{l}\Sl\bactN)}{\partial t}+\divec (\Phi \rho_{l} S_{l}\textbf{D}_{\bactN,l}\nabla \bactN )=(\psigrowth-\psidecay) \Phi \rho_{l}\Sl\bactN, 
\end{align}  
where $\Phi$ is the medium porosity, $D_{\bactN,l}$ the microbial diffusion coefficient ($\unit{\square\meter\per\second}$), $\rho_{l}$ the liquid phase density $(\unit{\kilogram\per \meter^3})$, and $\Sl$ the liquid saturation. The microbial population is normalized as $\bactN = n/n_0$, where $n$ (respectively $n_0$) denotes the current (respectively initial) microbial density $(\unit{\meter^{-3}})$.  A double Monod model (\cite{monod1949}) is employed for the growth rate function meanwhile the decay rate is formulated as a linear rate (\cite{Khoshnevis2023}):
 \label{Biorates}
  \begin{align}
    & \psigrowth =\psigrowthmax\left(\frac{\xHtwol}{\lHtwo+\xHtwol}\right)\left(\frac{\xCOtwol}{\lCOtwo+\xCOtwol}\right),\\
    & \psidecay = \bdecay\bactN =\bdecay\frac{n}{n_0},\\
  \end{align}
Here, $\xHtwol$ and $\xCOtwol$ denote the molar fractions of \ce{H2} and \ce{CO2}, respectively, in the liquid phase; $\lHtwo$ and $\lCOtwo$ are the corresponding half-saturation constants; $\psigrowthmax$ is the maximum growth rate $(\unit{\second^{-1}})$ and $\bdecay$  is the decay rate $(\unit{\second^{-1}})$. Several experimental values obtained essentially in laboratory are available in the literature (see Table\ref{tab:parameters-sources}).  Note that our results and implementations can be easily extended to other reactions involving \(\ce{H2}\) consumption, such as sulfide reduction~\cite{hogeweg2024development}.   Several experiments showed that the dimensionless form of the Monod model mitigates numerical instability caused by extreme microbial density values, ensuring better conditioning and scaling with the overall-composition mass-fraction formulation~\cite{SAFARI2024120426}. Although methanogenic archae can displace in the liquid phase due to chemotaxis (\cite{maier2009environmental}, \cite{Becker2014}), the transport of the microbial population is also neglected in the present work. 

\begin{table}[h!]
\centering
\scriptsize
\caption{Range of Parameters for Simulating Compositional Flow with Bacteria}
\label{tab:parameters-sources}
\begin{tabular}{l l l}
\toprule
\textbf{Parameter} & \textbf{Range} & \textbf{Reference} \\
\midrule
\(\psigrowthmax\) ($\unit{\second^{-1}}$) & $1.805 \times 10^{-6} - 3.009 \times 10^{-5}$  & \cite{Elferink1994,Elferink1995,Karadagli2005}\\
&&\cite{Robinson1984,Kristjansson1982}\\
&&\cite{Ebigbo2013,Stams2003,Vavilin2000}\\
\(\lHtwo\)(\unit{\mole/\mole}) &$9 \times 10^{-12} - 3.240 \times 10^{-7}$ & \cite{Elferink1994,Elferink1995,Karadagli2005} \\
&&\cite{Robinson1984,Kristjansson1982}\\
&&\cite{Stams2003,Ebigbo2013}\\
&&\cite{Dornseiﬀer1995,Vavilin2000}\\
\(\lCOtwo\) (\unit{\mole/\mole})&  $2.34 \times 10^{-8} - 5.4 \times 10^{-6}$ & \cite{Ebigbo2013,Dornseiﬀer1995,Vavilin2000} \\
 \(\YHtwo\) ($mol^{-1}$) &  $7.697 \times 10^{10} - 1.093 \times 10^{13}$ & \cite{Elferink1994,Elferink1995,Karadagli2005}  \\
 &&\cite{Robinson1984,Ebigbo2013,Vavilin2000}\\
\(\bdecay\) ($\unit{\second^{-1}}$) & $6.944 \times 10^{-7} - 1.019 \times 10^{-6}$ & \cite{Karadagli2005,Ebigbo2013} \\
\(T\) ($\unit{\kelvin}$) & $298.15 - 313.15$ & \cite{Vavilin2000,eddaoui2021,Kotsyurbenko2001} \\
 &&\cite{Panfilov2018}\\
\(P\) ($\unit{\bar}$) & $80 - 100$ & \cite{eddaoui2021,Panfilov2018} \\
\bottomrule
\end{tabular}
\end{table}

\subsection{Overall composition formulation}
\label{subsection:overallcomposition}
The microbial model is coupled to the two-phase compositional model for fluid flow through  a bioreactive source term in the mass conservation equations. This model consists of mass conservation equations for each component based on the overall composition formulation~\cite{VOSKOV2012101} and the multiphase extension of Darcy's law, which governs fluid phase velocity:  
\begin{align}
    \label{Main_problem_model_cons}
     \partial_{t}(\Phi \rhot\zi)+\divec \ui +\qi & = \Phi \gamma^{\ce{H2}}_{i}\dfrac{\psigrowth \bactN  \Sl
     }{\YHtwo} ,\quad \chemset{H2, H2O, ...}.
\end{align}
Here, $\rhot$ denotes the total mass density, $\zi$ the overall mole fraction of the component $i$, $\ui$ the flux, and $\qi$ represents external sources, such as well sources.  Note that the pressure and the overall molar composition are chosen as the primary variables. The microbial contribution is modeled through the source term $\Phi \gamma^{\ce{H2}}_{i} \dfrac{\psigrowth \bactN \Sl}{\YHtwo}$~\cite{Panfilov2018,eddaoui2021}, where $\YHtwo$ is the yield coefficient and  $\gamma^{\ce{H2}}_{i} = n_{0}\gamma_{i} \molmass_i / \gamma_{\ce{H2}}$ accounts for the stoichiometric conversion of hydrogen into other species $i$, with the stoichiometric coefficients in reaction~\eqref{bioreaction} given by $[\gamma_{\ce{H2}},\gamma_{\ce{CO2}},\gamma_{\ce{CH4}},\gamma_{\ce{H2O}}] = [-4,1,1,2]$. The overall molar fraction $\zi$ and the total density $\rhot$ are defined as:
\begin{subequations}
  \begin{align}
    & \zi = \dfrac{\sum_{\zeta\in\{l,g\}}\rhoalpha\salpha\Xalphai}{\sum_{\zeta\in\{l,g\}}\rhoalpha\salpha} = \sum_{\zeta\in\{l,g\}}\Aalpha\Xalphai,\\
    &  \rhot = \sum_{\zeta\in\{l,g\}}\rhoalpha\salpha.
  \end{align}
\end{subequations}
where $\Aalpha$ represents the contribution of each phase ($\zeta \in \{l,g\}$) to the total mass density $\rhot$, and $\ui$ is the species flux made up of two contributions: the standard multiphase extension of Darcy's law, $\valpha$, and the diffusive flux, $\jialpha$, of component $i$ in phase $\zeta$. These are given by:
\bse  
\begin{alignat}{2}
\label{Main_problem_model_velocity}  
  &\ui = \sum_{\zeta\in\{l,g\}}\rhoalpha\Xalphai\valpha -\jialpha,\\
&\valpha = -\frac{k_{r\zeta}}{\mu_\zeta}\vK(\nabla \palpha-\rhoalpha\mathbf{g} ), \quad \zeta\in\{\waterphase,\,\gasphase\},\\
\label{Main_problem_model_diffusion}      
&\jialpha = - \Phi\rhoalpha\Salpha\textbf{D}_{\zeta,i}\nabla \Xalphai, \quad \chemset{H2, H2O, ...}.
\end{alignat}

\ese Here, $\valpha$ is the phase velocity from the generalized Darcy's law, where $k_{r\zeta}$ is the relative
permeability, $\mu_\zeta$ is the viscosity, $\vK$ is the absolute permeability tensor, and $\palpha$ is the phase
pressure. The term $\jialpha$ represents molecular diffusion flux, driven by the gradient of the mole fraction $\Xalphai$, with $\textbf{D}_{\zeta,i}$ being the diffusion coefficient. Besides $\textbf{D}_{\zeta,i}$ depends on the tortuosity $\tau_\zeta$ and binary molecular diffusion coefficients $\textbf{D}_{\zeta,i}^*$ (see table \ref{tab:moleculardiffusioncoef}) such as $\textbf{D}_{\zeta,i}=\textbf{D}_{\zeta,i}^* \tau_{\zeta}$. The Millington and Quirk model is used for the tortuosity (~\cite{MillingtonQuirk1961}) and writes as:
\begin{align}
&\label{millingtonquirk}  
\tau_{\zeta}= \frac{1}{\Phi^2}(\Phi \Salpha)^{7/3}.
\end{align}
To fully determine the unknowns in the coupled system \eqref{eq:bactmodel}$-$\eqref{millingtonquirk}, we must compute the phase saturations and the mole fractions of all components in each phase. Moreover, since both phase densities and flow potentials may exhibit pressure dependence, knowledge of the individual phase pressures is required. Crucially, evaluating phase behavior and flow properties demands an accurate description of component distribution between the fluid phases. This is achieved by solving the following set of nonlinear closure relations:
\bse
\label{closure_model}  
\begin{align}
&\label{Main_problem_model_cons_sat}\sum_{ \zeta\in\{\waterphase,\,\gasphase\}}\salpha=\sum_{ \zeta\in\{\waterphase,\,\gasphase\}}\Aalpha =1,\\\label{Main_problem_model_fugacity}
& \sum_{i}^{N_c}\zi =  \sum_{i}^{N_c}\Xalphai=1,\quad \forall\zeta\in\{\waterphase,\,\gasphase\},\\
&\po -\pg-\pclg=0,\quad f_{l,i} -f_{g,i}=0, \quad \chemset{H2, H2O, ...},
\end{align}
\ese
Given that the equation of state is properly calibrated against experimental data, the compositional model can accurately capture phase behavior and interphase mass transfer processes. This is particularly important in the presence of microbial activity, where chemical transformations such as hydrogen consumption, carbon dioxide production, and biomass accumulation can significantly alter phase compositions, fluid properties, and ultimately reservoir performance. Note that these equations are coupled with well equations and specific initial and boundary conditions. For more details on the model, see~\cite{moyner2021compositional}. The compositional fluid flow model was solved using the MRST object-oriented automatic
differentiation framework \cite{krogstad2015mrst,mrst-book-2-ad}. By default, MRST employs a fully implicit two-point
flow approximation (TPFA) with a single-point upwinding to solve the flow equations. Stability and flash calculations are
performed at each Newton iteration for the compositional model.
       
\begin{table}[h!]
\centering
\scriptsize 
\caption{Binary molecular diffusion coefficients of components \ce{H2}, \ce{C1}, \ce{H2O}, \ce{CO2}, \ce{N2}, \ce{C2}, \ce{C3}, and \ce{NC4} in both liquid and gas phases~\cite{Wilke1955, Cussler2009}.}
\label{tab:moleculardiffusioncoef}
\begin{tabular}{l l l l l l l l l}
\toprule
& \ce{H2} & \ce{C1} & \ce{H2O} & \ce{CO2} & \ce{N2} & \ce{C2} & \ce{C3} & \ce{NC4} \\
\midrule
$\textbf{D}_{l,i}^*(\unit{\square\meter\per\second})$  
& $4.5 \times 10^{-9}$ & $2.6 \times 10^{-9}$ & $2.3 \times 10^{-9}$ & $1.9 \times 10^{-9}$ & $2.1 \times 10^{-9}$ & $1.2 \times 10^{-9}$ & $0.9 \times 10^{-9}$ & $0.7 \times 10^{-9}$ \\
$\textbf{D}_{g,i}^*(\unit{\square\meter\per\second})$ 
& $6.1 \times 10^{-5}$ & $2.5 \times 10^{-5}$ & $1.5 \times 10^{-5}$ & $1.4 \times 10^{-5}$ & $1.8 \times 10^{-5}$ & $1.6 \times 10^{-5}$ & $1.2 \times 10^{-5}$ & $0.9 \times 10^{-5}$ \\
\bottomrule
\end{tabular}
\end{table}

\subsection{Impact of microbial activity on rock properties}
When methanogenic archae grow, they obstruct the pore space affecting the fluid flow through the porous medium (\cite{Eddaoui_Panfilov2021}). Hence, the porosity and the permeability are impacted. The porosity $\Phi$ and permeability $\vK$ vary with bacteria density $\bactN$ as:
\begin{align}
  \label{eq:bioclogging}
  \Phi(\bactN) &= \frac{\Phi_0}{1+ c_p (\bactN/\bactN_c)^2}, \\
  \vK(\bactN) &= \vK_0\left(\frac{1-\Phi_0}{1-\Phi}\right)^2 \left(\frac{\Phi}{\Phi_0}\right)^3,
\end{align}
where $\bactN_c$ is the critical density for maximal clogging, $\Phi_0$ and $\vK_0$ are reference values, and $c_p$ controls porosity reduction (\cite{Eddaoui_Panfilov2021}).

\section{Thermodynamic models}\label{section:thermomodels}

In this section, we aim at employing a thermodynamic model able to take into account the solubility of gases such as hydrogen and carbon dioxide in two-phase and multi-component problems. An accurate estimation of the solubility of gases reveals to be essential to predict the Vapor-Liquid Equilibrium ($VLE$) and the phase compositions of the multi-component system.

\subsection{EoS models adapted to capture gas solubility}

In an isothermal and isobaric system, the coexistence of several phases results in the equality of chemical potentials of
each component in all phases with respect to the Gibbs-Duhem equation (\cite{Gibbs1948}). For $VLE$, the equality of chemical potentials is expressed as the equality of fugacities~\eqref{Main_problem_model_fugacity} \cite{Helmig1997}.

In the literature, two approaches are available to calculate $VLE$: the symmetric and asymmetric approach. In the symmetric method, the fugacity coefficients of each component on both phases are calculated from a cubic (e.g. Peng-Robinson (PR), Soave-Redlich-Kwong (SRK), etc.) or a non-cubic (e.g. SAFT, CPA, multiparameter Helmholtz energy,..., -type) equation of state (EoS). In the asymmetric approach, the fugacity coefficients of each component in the vapor phase are calculated as in the symmetric approach, while those of the liquid phase are determined using the model $G-\text{excess}$ ~\cite{Lee2021}.\\
Futhermore, in their work on the VLE of \ce{H2}-\ce{H2O} mixtures, \cite{Rahbari2019} have proved that using cubic EoS, such as PR or SRK, with classical mixing rules are unable to accurately predict the coexistence composition of the non aqueous and aqueous phases, even if adapted Binary Interactions Parameters (BIPs) were introduced.\\
However, ~\cite{soreide1992peng} have elaborated a model specifically tailored for brine solutions which consists in adding salt molality and temperature dependencies in the alpha-function of the PR-EoS and the BIPs between water and gases in the aqueous phase. Then, the model has been improved by Chabab et al. for \ce{CO2} and \ce{H2} solubilities in water and NaCl brine.(\cite{chabab2019thermodynamic, CHABAB2024648}).\\
To fully describe the thermodynamic behavior of the multicomponent fluid, our compositional model implemented in MRST must be complemented with an appropriate equation of state. The fluid is assumed to consist of brine, treated as a pseudo-component representing \ce{H2O} and \ce{NaCl}, along with $n-1$ additional light components, such as \ce{CO2}, \ce{N2}, and \ce{CH4}. In this work, \ce{NaCl} is chosen as the representative salt. 

To model the phase equilibrium of this multi-component system, we have implemented the enhanced version of the SW EoS found in the literature (\cite{chabab2020measurements,CHABAB2024648}). The PR EoS is expressed as \cite{chabab2020measurements}:

\begin{equation}
    \press(\temp,\molarvol) = \frac{\gasconst \temp}{\molarvol - \eoscoeffb} - \frac{\eoscoeffa}{\molarvol (\molarvol + \eoscoeffb) + \eoscoeffb (\molarvol - \eoscoeffb)},
    \label{eq:7}
\end{equation}
where $\gasconst$ is the universal gas constant, $\temp$ is the temperature and $\molarvol$ is the molar volume, $\eoscoeffa$ and $\eoscoeffb$ are the EoS parameters. \\
In a mixture, $\eoscoeffa$ and $\eoscoeffb$ are the EoS parameters of the mixture and depend on the temperature, the molar fraction and the phase. They are noted $\eoscoeffa_{\zeta}(\temp, \comp)$ and $\eoscoeffb(\comp)$ and are determined using standard mixing rules, namely:
\begin{align}
    \eoscoeffa_{\zeta}(\temp, \comp) = \sum_{i,j=1}^{Nc} \sqrt{\eoscoeffa_{i}(\temp) \eoscoeffa_{j}(\temp)} (1 - \BIP_{\zeta,ij}) \Xalphai \Xalphaj,\quad\eoscoeffb(\comp) = \sum_{i=1}^{Nc} \eoscoeffb_{i} \Xalphai, \label{eq:8}
\end{align}
where 
\begin{align}
   & \eoscoeffa_{i}(\temp) = \Omega_a \frac{\gasconst^2 T_{c,i}}{P_{c,i}} \eosalpha_{i}(\temp), \quad  \eoscoeffb_{i} = \Omega_b \frac{\gasconst T_{c,i}}{P_{c,i}},\label{eq:8d}
\end{align}
with $\Omega_a = 0.45724$, $\Omega_b = 0.0778$ constants. $P_{c,i}$ and $T_{c,i}$ represent the critical pressure and temperature for component $i$, respectively.

In both gaseous and liquid phases, we use the generalized alpha function, $\eosalpha_{i}$, proposed by Soave (\cite{Soave1972})  and  formulated as:
 
\begin{align}
 &\sqrt{\eosalpha_{i}} = 1 + (0.37464 + 1.54226 \acf_{i} - 0.26992 \acf_{i}^2) \left(1 - \sqrt{T_{r,i}}\right).
\label{eq:9a}
\end{align}
Here $T_{r,i} = T/T_{c,i}$ is reduced temperature corresponding to the critical temperature $T_{c,i}$ and $\acf_{i}$ is the acentric factor for component $i$ 

The first modification applied in the SW model  (\cite{soreide1992peng}) consists in including a salt molality, $\msalt$ ($\unit{\mole\per\kgw} $), dependence  for the component \ce{H2O} in the liquid phase, namely:
\begin{align}
&\sqrt{\eosalpha_{\ce{H2O}}} = 1 + 0.453 (1 -T_{r,\ce{H2O}}(1-0.0103\msalt^{1.1})) + 0.0034 (T_{r,\ce{H2O}}^{-3}-1).
\label{eq:9b}
\end{align}

The second modification in the SW model consists in adding particular (phase-specific) BIPs, $\BIP_{\zeta,ij}$ , in the mixing rules in order to represent the cross interactions between the $i$th and $j$th components and more particularly between component \ce{H2O} and the others in each phase. The BIPs are symmetric, that is $\BIP_{\zeta,ij}=\BIP_{\zeta,ji}$. Given that the thermodynamical conditions in UHS reveal to be far from the \ce{H2O} critical point, the use of phase-specific BIPs remains suitable (\cite{CHABAB2024648}).

We compile a comprehensive dataset from the literature for BIPs across a wide range of mixtures. As an illustrative example, the BIPs formulations can be unified in the following generalized forms:
\begin{align}
&\BIP_{l,\ce{CO2}-\ce{H2O}}  = T_{r,\ce{CO2}} (a_1 + a_2 T_r + a_3 T_{r,\ce{CO2}} \msalt) + \msalt^2 (b_1 + b_2 T_{r,\ce{CO2}}) + b_3, \nonumber\\
&\BIP_{l,j-\ce{H2O}}  =  a_1 (1 + b_1 \msalt^{c}) + a_2 (1 + b_2 \msalt^{c}) T_{r,j} + a_3 (1 + b_3 \msalt^{c})T_{r,j}^2+ a_4 \exp{(b_4 T_{r,j})}, \nonumber\\
&\BIP_{g,j-\ce{H2O}} = a_1 T_{r,j} + a_2.
\label{eq:bic_coef}
\end{align}
As an example, for the mixture $[\ce{H2},\ce{H2S},\ce{N2}, \ce{C1}, \ce{C2}, \ce{C3}, \ce{NC4}]$, the corresponding coefficients are provided in Table~\ref{tab:coefficientsliqgas}.
\begin{table}[h!]
\centering
\scriptsize 
\caption{Coefficients for generalized BIPs equations in liquid/gas phases \cite{AFANASYEV2021,ChababCruz2021,CHABAB2024648}}
\label{tab:coefficientsliqgas}
\begin{tabular}{l l l l l}
\toprule
&&\qquad\textbf{liquid-phase}&&\\
\toprule
 & \textbf{\ce{CO2}} & \textbf{\ce{H2}} & \textbf{\ce{C1}} & \textbf{\ce{H2}S} \\
\midrule
$a_1$ & $-0.43575155$ & $-2.11917$ & $-1.625685$ & $-0.42619$ \\
$a_2$ & $-0.05766906744$ & $0.14888$ & $1.114873$ & $0.673586$ \\
$a_3$ & $0.00826464849$ & - & - & - \\
$a_4$ & - & $-13.01835$ & - & - \\
$b_1$ & $-0.00129539193$ & $-0.0226322$ & $8.590105 \times 10^{-21}$ & $-0.0575$ \\
$b_2$ & $-0.0016698848$ & $-0.0044736$ & $0.001812763$ & $-0.078823343$ \\
$b_3$ & - & - & $-0.169968$ & $-0.21625$ \\
$b_4$ & - & $-0.43946$ & $-0.04198569$ & $-0.160085087$ \\
$c$ & $-0.47866096$ & $1$ & $1$ & $1$ \\
\bottomrule
\end{tabular}
\\
\vspace{0.5cm}
\centering
\scriptsize 
\begin{tabular}{l l l l l}
\toprule
&&\qquad\textbf{liquid-phase (Cont.)}&&\\
\toprule
& \textbf{\ce{N2}} & \textbf{\ce{C2}} & \textbf{\ce{C3}} & \textbf{\ce{NC4}} \\
\midrule
$a_1$ & $-1.702359096$ & $1.1120 - 1.7369 \omega^{0.1}$ & $1.1120 - 1.7369 \omega^{0.1}$ & $1.1120 - 1.7369 \omega^{0.1}$ \\
$a_2$ & $0.450487$ & $1.1001 + 0.8360 \omega$ & $1.1001 + 0.8360 \omega$ & $1.1001 + 0.8360 \omega$ \\
$a_3$ & - & - & - & - \\
$a_4$ & - & - & - & - \\
$b_1$ & $0.01792130$ & $0.017407$ & $0.017407$ & $0.017407$ \\
$b_2$ & $0.066426$ & $0.033516$ & $0.033516$ & $0.033516$ \\
$b_3$ & - & $-0.15742 - 1.0988 \omega$ & $-0.15742 - 1.0988 \omega$ & $-0.15742 - 1.0988 \omega$ \\
$b_4$ & - & $0.011478$ & $0.011478$ & $0.011478$ \\
$c$ & $0.8$ & $1$ & $1$ & $1$ \\
\bottomrule
\end{tabular}
\\
\vspace{0.5cm}
\centering
\scriptsize 
\begin{tabular}{l l l }
\toprule
&\qquad\textbf{gas-phase}&\\
\toprule
& $a_1$ & $a_2$ \\
\midrule
{\ce{CO2}}&$0.207440935$&$-2.066623464504 \times 10^{-2}$ \\
 {\ce{H2}}&$0.042834$& $0.01993$\\
 {\ce{C1}}&-&$0.494435$ \\
\ce{H2S}&$-0.05965$&$0.19031$ \\
\ce{N2}&-&$0.385438$ \\
\ce{C2}&-&$0.4920$ \\
\ce{C3} &-&$0.5525$ \\
\ce{NC4} &-&$0.5091$ \\
\bottomrule
\end{tabular}
\end{table}
Like PR EoS, the input parameters for  each fluid component are critical properties ($\Pc$, $\Tc$), acentric factor ($\acf$), and molar mass ($\molmass$). These properties are essential for accurately modeling the phase behavior of the multicomponent system. The critical properties and acentric factors for the components considered in this study come from the \href{http://www.coolprop.org/}{Coolprop library} and are listed in Table~\ref{tab:component_properties}.

{Concerning the implementation of SW EoS in MRST, for reasons of simplicity, salinity in SW EoS has been considered as a constant input parameter and not as a composition that can vary (e.g. through water evaporation) as part of the stability and flash calculations. For a two-phase simulation and at conditions far from the critical point of the gas/water mixture, this assumption is fairly acceptable \cite{petitfrere2016three}. It is worth noting that recently, methods have been developed which consider salinity in SW EoS as a composition (and not a parameter) and include it in equilibrium calculations \cite{petitfrere2016three} \cite{panfili2024implementation}. These modifications, which make the SW model thermodynamically consistent, are especially necessary for applications with more than two phases, at very high temperatures (close to the critical point of water), and when trapping by dissolution is the dominant mechanism (e.g. \ce{CO2} sequestration).}

\begin{table}[!ht]
\centering
\scriptsize 
\caption{Components properties from the Coolprop library.}
\label{tab:component_properties}
\begin{tabular}{l l l l l l}
\toprule
Component & $\Pc (\unit{\pascal})$ & $\Tc (K)$ & $V_c (\unit{\meter^3})$ &  $\acf$ & $\molmass (\unit{\kilogram \mole^{-1}})$ \\
\midrule
\ce{H2O} &$2.21\times10^{7}$ & $647.1$ &  $5.595\times10^{-5}$ & $0.344$ & $0.0180153$ \\
\ce{H2} &$1.30\times10^{6}$ & $33.1$ &  $6.448\times10^{-5}$ & $-0.219$ & $0.0020159$ \\
\ce{CO2} &$7.38\times10^{6}$ & $304.1$ &  $9.412\times10^{-5}$ & $0.224$ & $0.0440098$ \\
\ce{CH4}/\ce{C1} &$4.60\times10^{6}$ & $190.6$ &  $9.863\times10^{-5}$ & $0.011$ & $0.0160428$\\
\ce{C2} &$4.87\times10^{6}$ & $305.3$ &  $1.458\times10^{-4}$ & $0.099$ & $0.0300690$ \\
\ce{C3} &$4.26\times10^{6}$ & $369.8$ &  $9.418\times10^{-5}$ & $0.152$ & $0.0440962$ \\
\ce{NC4} &$3.80\times10^{6}$ & $425.1$ &  $2.549\times10^{-4}$ & $0.201$ & $0.0581222$ \\
\ce{N2} &$3.40\times10^{6}$ & $126.2$ &  $8.941\times10^{-5}$ & $0.037$ & $0.0280135$ \\
\bottomrule
\end{tabular}
\end{table}

\subsection{Hydrogen and carbon-dioxide solubility results: PR vs SW EoS}
\label{subsection:SW_EoS}
In MRST the VLE problem is solved using a symmetric approach with a PR or SRK Equation of State. Then, the obtained
Flash equations are solved as a nested non linear system (\cite{mrst-book-2-ad}).  As previously mentioned, in this work, the Søreide-Whitson model has been implemented in MRST to account for the solubility of \ce{H2} and \ce{CO2} in water and brine.

Then, several numerical simulations have been performed on two-component $\ce{H2}-\ce{H2O}$ and $\ce{CO2}-\ce{H2O}$ and
two-phase systems to verify the efficiency of the SW model developed in MRST over a wide range of thermodynamical
conditions, i.e.: $100$ \unit{\bar} $\leq P \leq 200.11$ bar and
$298.15\,\unit{\kelvin}\leq T \leq 373.85\,\unit{\kelvin}$ for the $\ce{H2}-\ce{H2O}$ two-phase system and $17$ bar
$\leq P \leq 141.1$ bar and $323.2 \,\unit{\kelvin}\leq T \leq 373.41\,\unit{\kelvin}$ for the $\ce{CO2}-\ce{H2O}$
two-phase system. Then, they are compared with the experimental data provided by
\cite{chabab2020measurements,chabab2019thermodynamic}.

\begin{figure}
\centering
\includegraphics[width=0.5\linewidth]{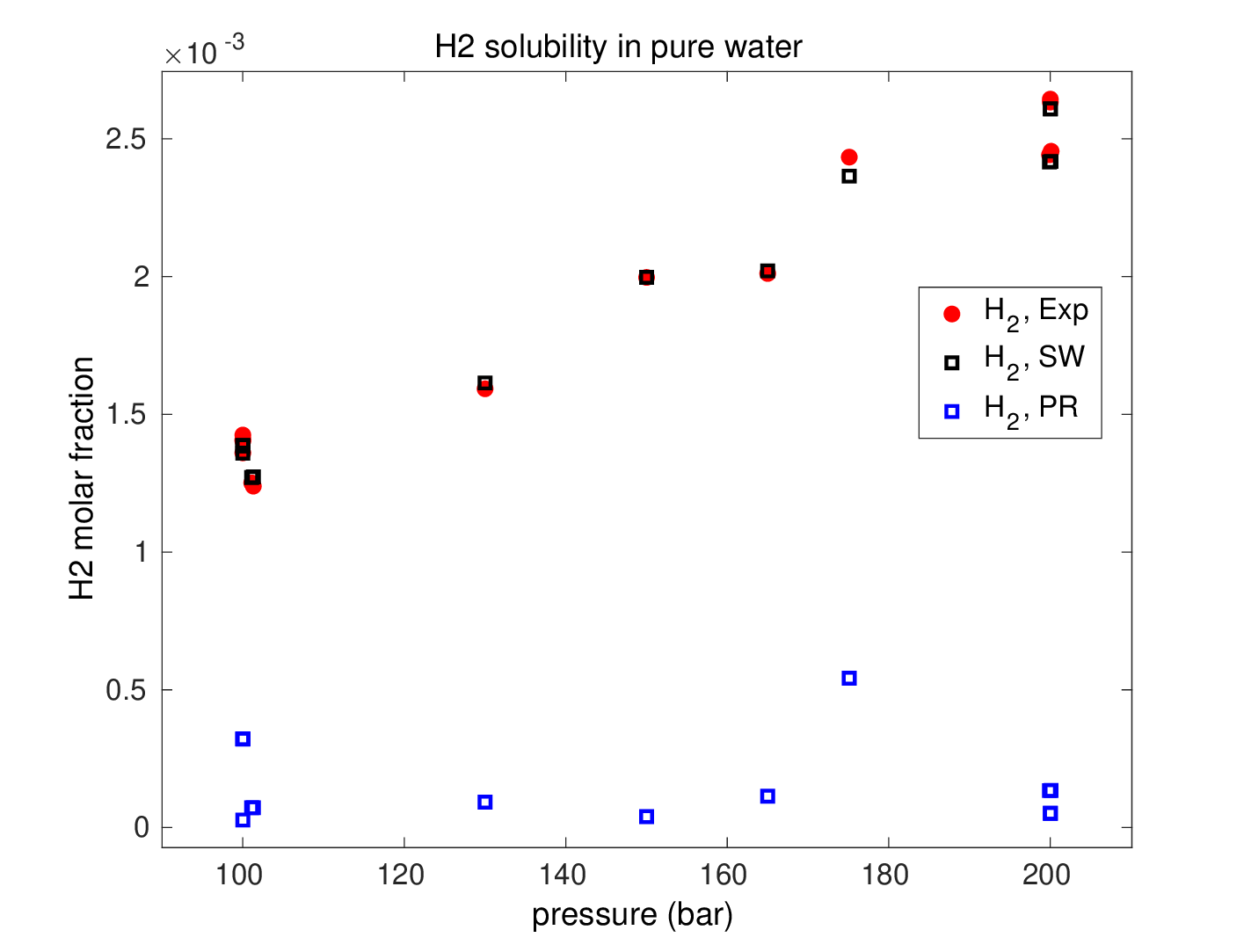}
 \includegraphics[width=0.48\linewidth]{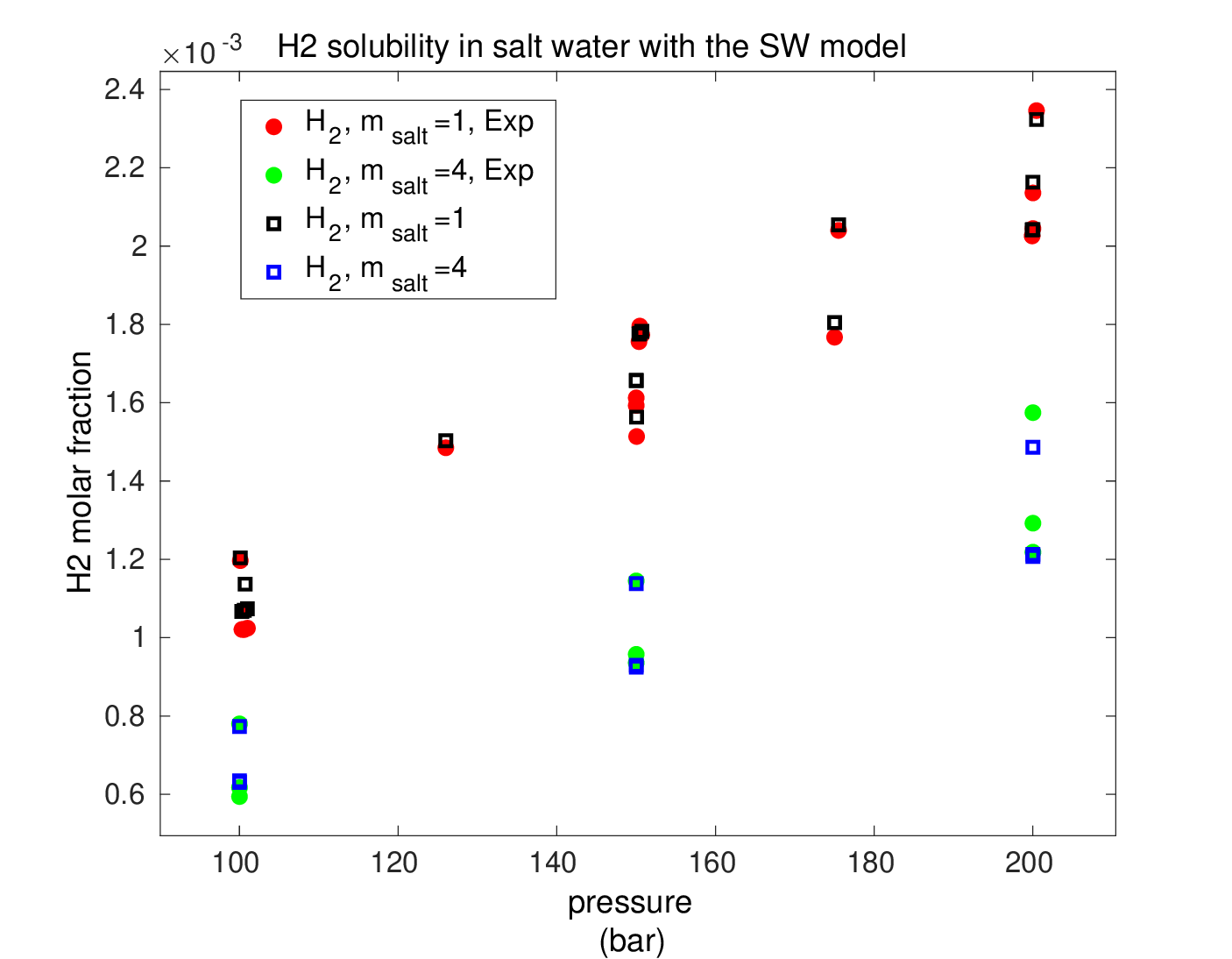}
\caption{\ce{H2} solubility as a function of pressure at different temperatures. Solubility in pure water is modeled using PR and SW equations of state (left), while solubility in saline water is shown using the SW model (right). Data collected from~\cite{CHABAB2024648}.}
\label{fig:H2SW}
\end{figure}

\begin{figure}
\centering
\includegraphics[width=0.5\linewidth]{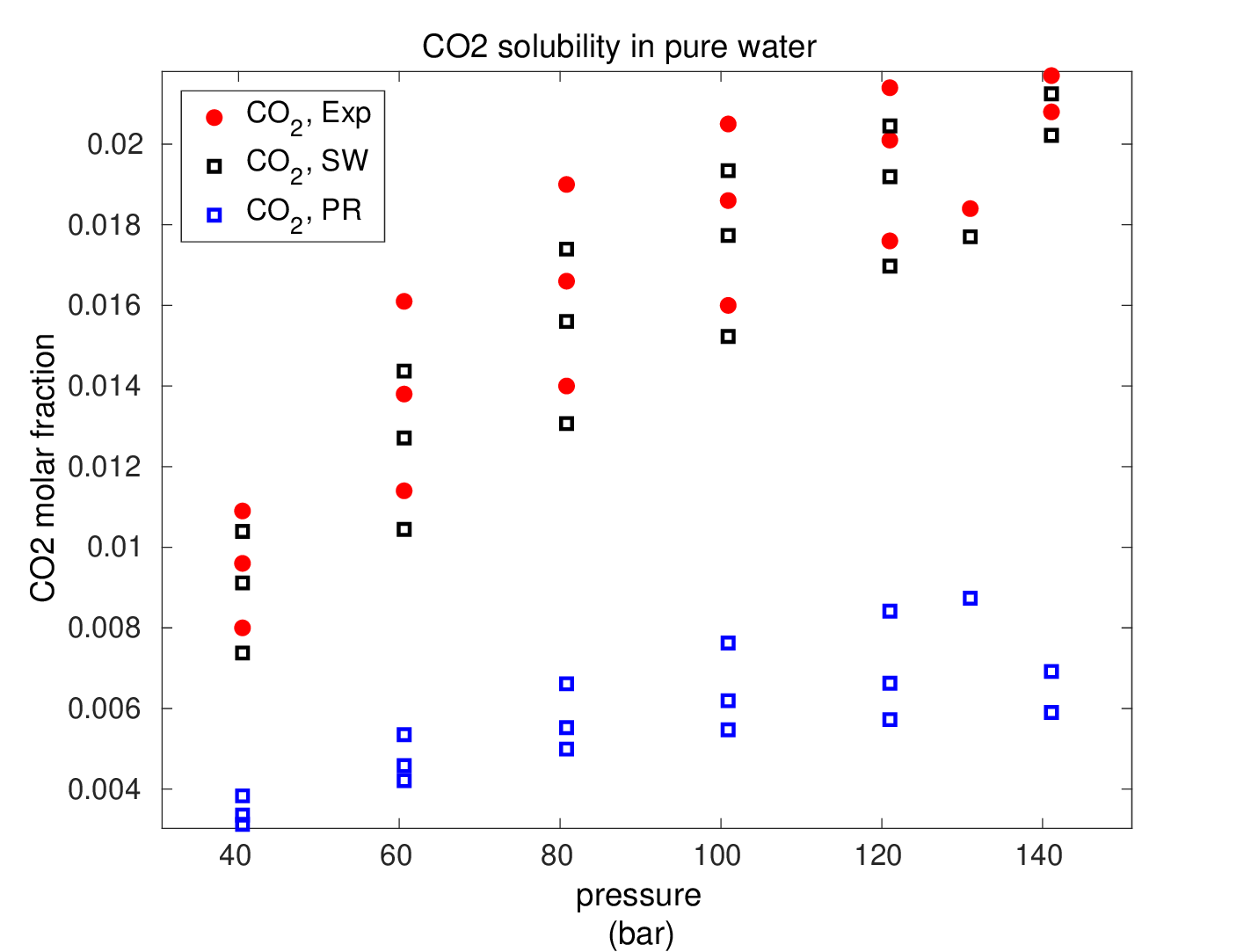} \includegraphics[width=0.48\linewidth]{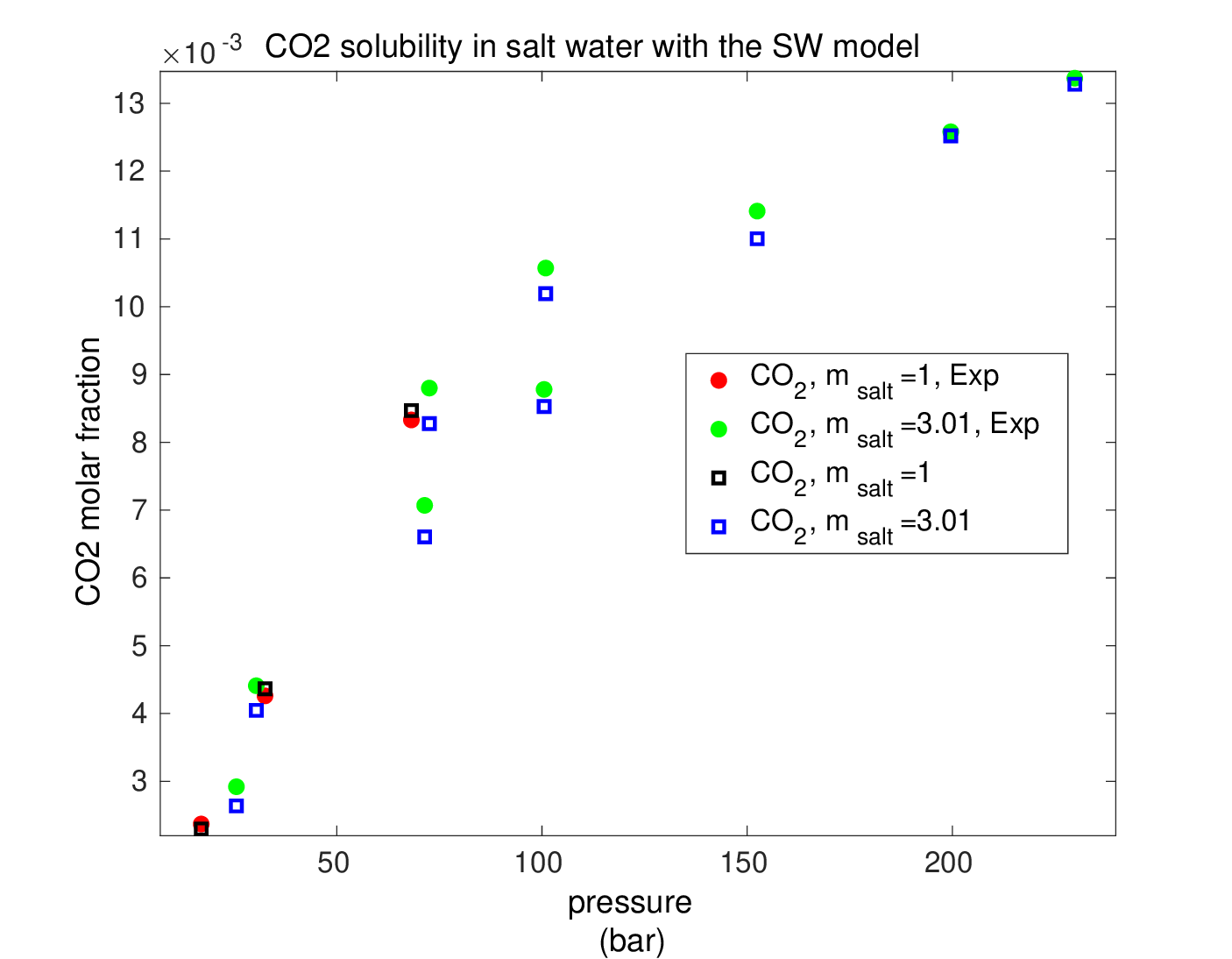}
\caption{\ce{CO2} solubility  along with pressure for different temperatures.\ce{CO2} solubility in pure captured with the PR and SW EoS models (left). \ce{CO2} solubility in salt water captured with the SW model (right). Data collected from~\cite{chabab2020measurements}.}
\label{fig:CO2SW}
\end{figure}

As demonstrated in ~\cite{Rahbari2019}, Figure ~\ref{fig:H2SW} (left) shows the limits of the PR model to capture the \ce{H2} solubility in water whereas the SW model matches very well with the experimental data in water and brine, see
Figure \ref{fig:H2SW} (right). Relative errors are computed pointwise as $|X_{l,i} -X_{l,i}^{\text{exp}}|/|X_{l,i}^{\text{exp}}|$, where $i$ indexes the component $i$ and $l$ denotes the liquid phase. We report the minimum, mean, and maximum values over all points, see Table \ref{tab:errorsPRSW}. Therein, the PR model exhibits a mean error of $90.5\%$ (min error $77.1\%$) with the experiment data whereas the SW model does not exceed an error of $2.8\%$. Furthermore, it has to be noted that the solubility of \ce{H2} decreases when the salinity
increases. Similar results have been obtained for solubility of \ce{CO2} in water and brine, see Figure~\ref{fig:CO2SW}.

\begin{table}[h!]
\centering
\scriptsize 
\caption{Calculated errors between the EoS models and the experiment data.}
\label{tab:errorsPRSW}
\begin{tabular}{l l l l l l}
\toprule
EoS Model & Component Solubility & Salinity ($\unit{\mole\per\kgw}$) & $\text{error}_{\text{min}}$ & $\text{error}_{\text{mean}}$ & $\text{error}_{\text{max}}$ \\
\midrule
PR &\ce{H2} & 0 &  $0.7710$ & $0.9054$ & $0.9805$ \\
SW &\ce{H2} & 0 &  $0.0002$ & $0.0138$ & $0.0284$ \\
SW &\ce{H2} & 1 &  $0.0013$ & $0.0227$ & $0.0622$ \\
SW &\ce{H2} & 4 &  $0.0046$ & $0.0301$ & $0.0678$ \\
\midrule
PR  & \ce{CO2} & 0 & $0.5209$ & $0.6401$ & $0.7385$  \\
SW & \ce{CO2} & 0& $0.0208$ & $0.0566$ & $0.1074$ \\
SW &\ce{CO2} & 1 &  $0.0166$ & $0.0239$ & $0.0306$ \\
SW &\ce{CO2} & $3$ &  $0.0048$ & $0.0463$ & $0.0969$ \\
\bottomrule
\end{tabular}
\end{table}

\subsection{Compositional validation}
The compositional MRST model was validated against commercial solvers using the Peng-Robinson EoS. The test case is a 1D
problem from \cite{VOSKOV2012101}, later verified in \cite{mrst-book-2-ad}, injects pure \ce{CO2} (500 steps) into a
1,000-cell reservoir initially containing a \ce{CO2}-\ce{CH4}-\ce{C10H22} mixture (liquid-dominated, with supercritical
\unit{CO2} at 423.15 \unit{\kelvin}, 75\,\unit{\bar}). A similar validation using the SW-EoS (not shown) confirmed
agreement with MRST-PR and other commercial simulators.

\section{Numerical  simulations}\label{section:numericalresults}
In this work, we implemented the SW equation of state (see Section~\ref{section:thermomodels}) in MRST to accurately capture gas dissolution effects. This thermodynamic model allows for a consistent coupling with microbial dynamics, enabling the analysis of their impact on hydrogen loss, transport behavior, and bio-clogging mechanisms.

These developments are built on the compositional module of MRST (Lie, 2019), which provides advanced simulation capabilities for multicomponent systems. The module supports integration with external, pre-compiled linear solvers such as AMGCL (\cite{lie2019introduction,moyner2021compositional}). Through the use of memory-efficient backends for linearization and assembly, combined with high-performance solvers written in compiled languages, MRST can efficiently simulate both synthetic and field-scale reservoir models involving millions of unknowns.

The numerical results are structured as follows. In Subsection~\ref{example1D}, we demonstrate the effects of microbial
activity on \ce{H2} loss and methanation using a simple 1D model without wells, focusing on reservoir pressure,
saturation, and \ce{H2} dissolution. In Subsection~\ref{example2D}, we apply our fluid model to a 2D dome-shaped model with
different rock types and heterogeneous entry pressures. This section showcases the capability of our module to handle a
highly nonlinear model due to bio-clogging effects, varying capillary pressure, and different residual saturations, as
well as the impact of bacteria on \ce{H2} impurity. Subsection~\ref{Benchmark2023} demonstrates a 3D test case,
highlighting the interaction between salinity effects, the molecular diffusion and microbial activity, and their influence on \ce{H2} loss and production efficiency. Finally, in Subsection~\ref{Benchmark2022}, we test our model on a large UHS 3D test case with a complex mixture and realistic injection scenarios.

\subsection{Test 1:  1D compositional simulation}\label{example1D}

The simulation is conducted in a 1D reservoir domain, inspired by~\cite{moyner2018mass}, consisting of 1000 cells with a total length of $100$~m. This setup represents a simplified aquifer system where \ce{H2} and \ce{CO2} are present to study microbial activity and its impact on hydrogen loss. No wells are included in the simulation. We considered two initial compositions:
\begin{itemize}
    \item \textbf{Gas-rich case}: 40\% water, 44.5\% \ce{H2}, 5.5\% \ce{CH4}, and 10\% \ce{CO2}.  
    \item \textbf{Liquid-rich case}: 95\% water, 4\% \ce{H2}, 0.5\% \ce{CH4}, and 0.5\% \ce{CO2}.  
\end{itemize}
\begin{table}[h!]
\centering
\scriptsize
\caption{Microbial model parameters used in the MRST simulations}
\label{tab:microbial_parameters}
\begin{tabular}{l l l l l l l l}
\toprule
\textbf{Parameter} &\(\psigrowthmax\) (\(\si{\per\second}\)) & \(\bdecay\) (\(\si{\per\second}\)) &\(\YHtwo\) (\(\si{\per\mole}\)) & \(\lHtwo\) (-) & \(\lCOtwo\) (-)&\(n_0\) (\(\si{\per\cubic\meter}\))\\
\midrule
\textbf{Value} &\(\num{1.338e-4}\)  &\(\num{1.35e-6}\)  &\(\num{3.9e11}\)   &  \(\num{3.6e-7}\)& \(\num{1.98e-6}\) & \(\num{1.0e9}\)  \\ 
\bottomrule
\end{tabular}
\end{table}

The temperature is set at $313.75$~\unit{\kelvin}, and the initial pressure is 82~\unit{\bar}. The reservoir has a uniform initial permeability of $10$~\unit{\milli\darcy} and an initial porosity of 0.25, except in the final cells, where it is set to 0.9. A residual saturation is chosen for each phase $(S_{lr}, S_{or},S_{gr})=(0.2,0.2,0.1)$. Quadratic relative permeability functions are used for water and gas, with residual saturations \( S_{\mathrm{lr}} = 0.2 \) and \( S_{\mathrm{gr}} = 0.1 \). Effective permeabilities are obtained by shifting saturations and reinterpolating over the intervals \( [S_{\mathrm{lr}}, 0.9] \) for water and \( [S_{\mathrm{gr}}, 0.8] \) for gas. It should be noted that molecular and microbial diffusion are neglected. For the porosity reduction model, we set $c_p=1$ and $\bactN_{c} = 2n_0=180$. The densities of liquid and gas are set to their values under the surface conditions ($P=1.013\,\unit{\bar}$ and $T=15.56\,\unit{\celsius}$), that is, $\rho_l=999.014\; \unit{\kilogram\per\meter^3}$ and $\rho_g=0.084 \;\unit{\kilogram\per\meter^3}$.

The first simulation is abiotic; the second includes microbial effects but no clogging; and the third includes microbial clogging effects. Both simulations are performed with the SW EoS. The parameters for the microbial model are given in Table \ref{tab:microbial_parameters}.

\begin{figure}
\centering
\includegraphics[width=0.485\linewidth]{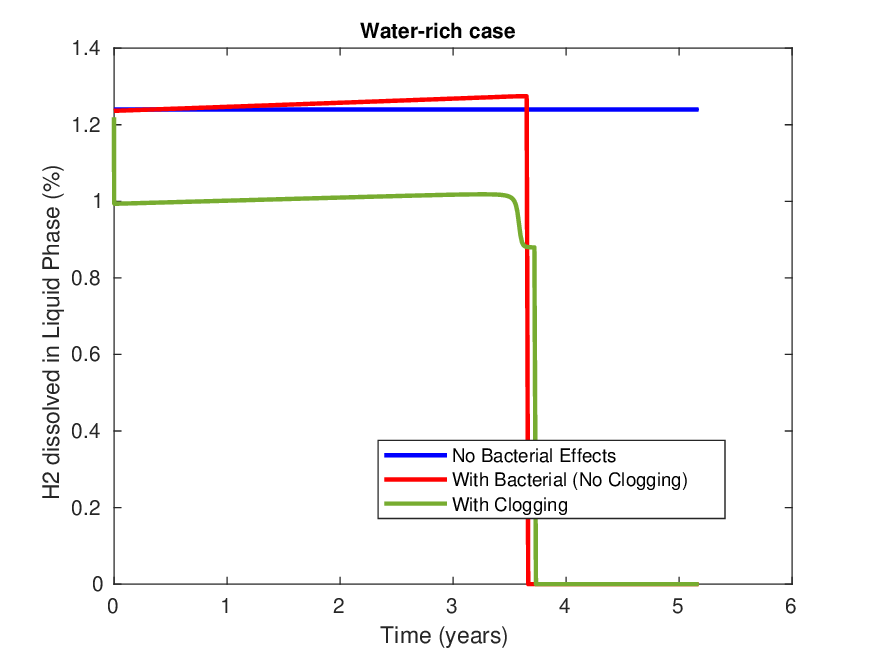}
\includegraphics[width=0.485\linewidth]{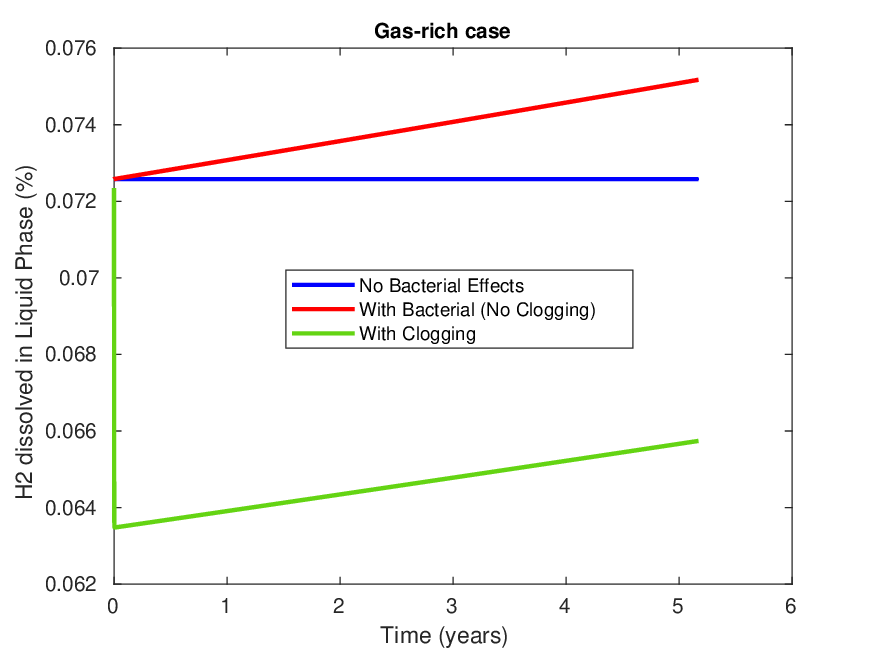}
\includegraphics[width=0.485\linewidth]{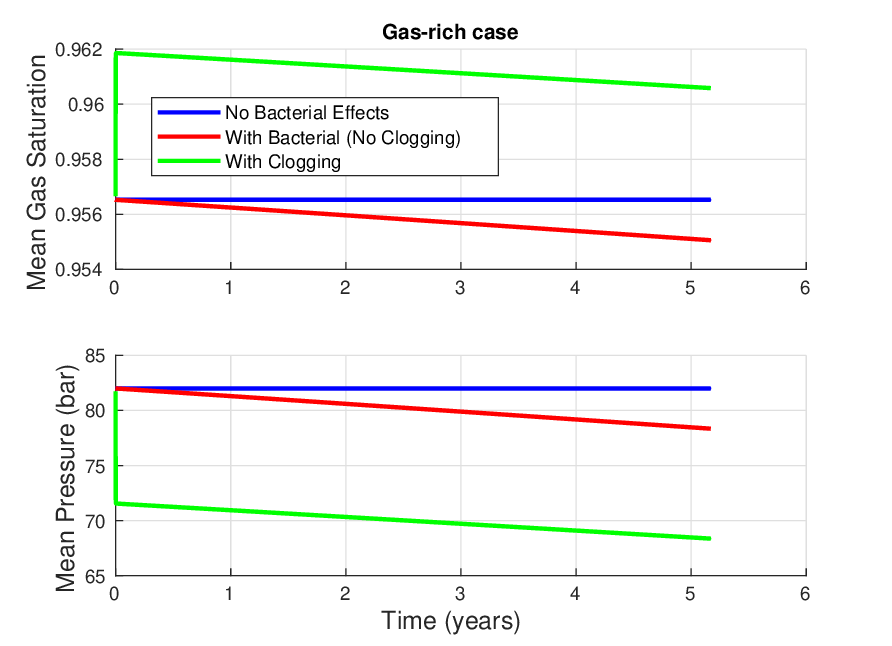}
\includegraphics[width=0.485\linewidth]{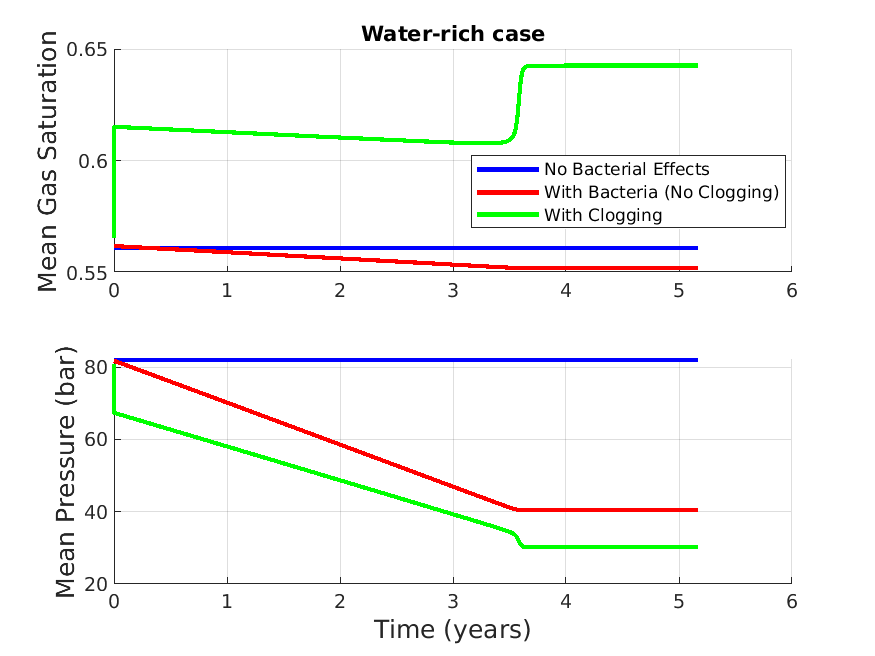}
\caption{1D case: Evolution of molar fractions at three times (top) and mean pressure and saturation (bottom).}
\label{fig:1d_mole_fractions}
\end{figure}

In Figure ~\ref{fig:1d_mole_fractions} (top), we  plot the dissolved \ce{H2} in the liquid phase. The values for the water-rich case are close to those reported in the literature, around 1.22\%, while for the gas-rich initialization,
they are significantly lower, around 0.072\%.  Clogging reduces dissolution in both cases but does not impact the total \ce{H2} consumption at the end of the simulation. The effects of bioclogging in both cases are observed primarily in the behavior of pressure and saturation, see Figure~\ref{fig:1d_mole_fractions} (top), and not in total \ce{H2} consumption. This is expected with a case without wells. The mean gas saturation decreases as \ce{H2} and \ce{CO2} are
consumed but increases with bioclogging, as reduced pore space and fluid flow trap gas. The pressure declines are clear: 5\%
(gas-rich) and 50\% (water-rich) without clogging, worsening to 17\% and 62\%, respectively, with clogging.

\begin{figure}
\centering
\includegraphics[width=0.485\linewidth]{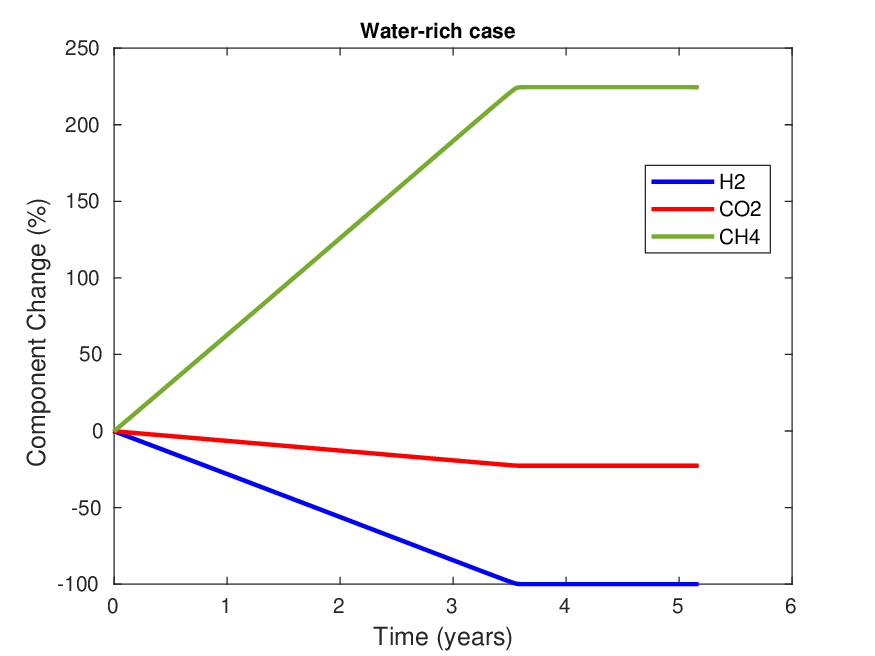}
\includegraphics[width=0.485\linewidth]{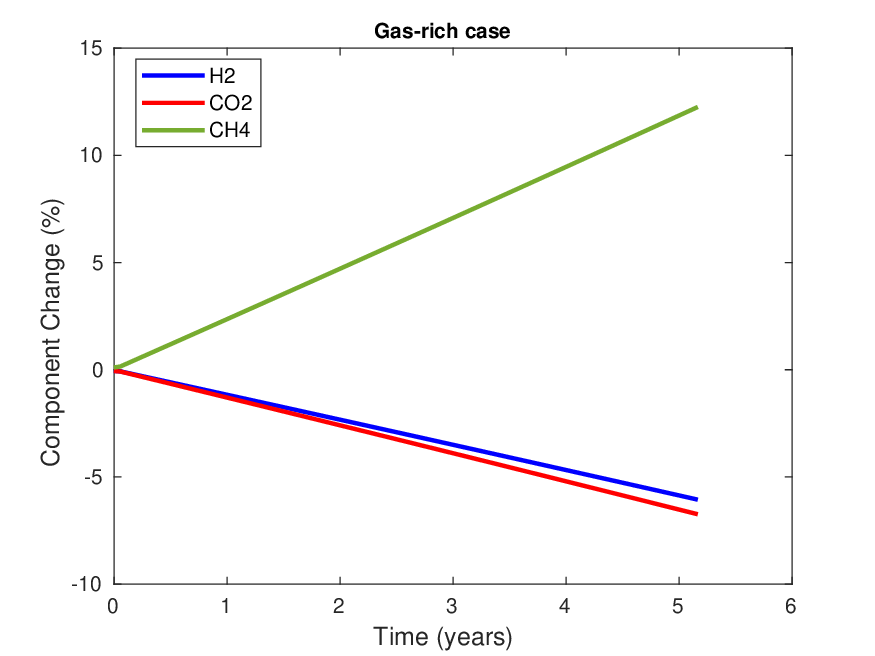}
\includegraphics[width=0.485\linewidth]{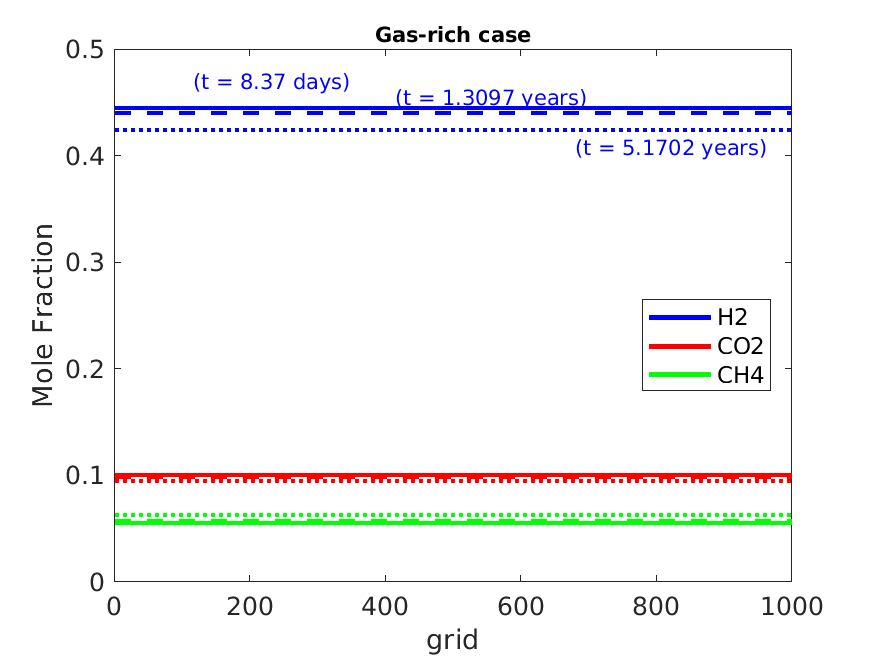}
\includegraphics[width=0.485\linewidth]{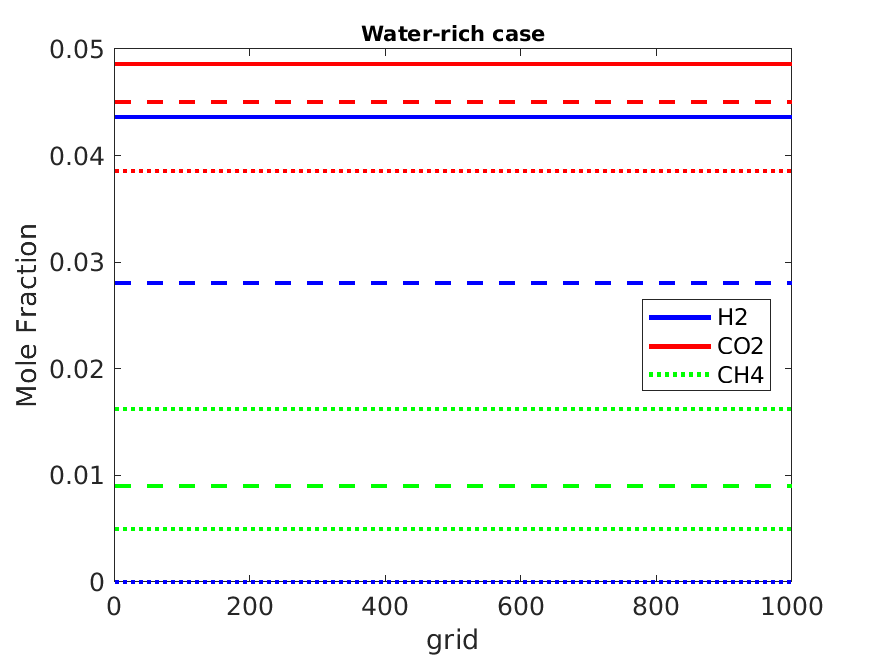}
\caption{1D case: \ce{H2} and \ce{CO2} loss vs. \ce{CH4} production (\%) (top) and dissolved \ce{H2} in the liquid phase (bottom), given by the ratio of liquid phase \ce{H2} mass to the total \ce{H2}  mass.}
\label{fig:1d_loss}
\end{figure}

In Figure~\ref{fig:1d_loss} (top), we plot the changes in the mixture due to microbial activity. As expected, microbial effects start immediately since \ce{H2} and \ce{CO2} are initially present in both gas-rich and water-rich cases. In the gas-rich case, the total consumption of \ce{H2} after 5 years of simulation is around 6.2\%.  In the water-rich one, all \ce{H2} was consumed after 3.5 years. Figure ~\ref{fig:1d_loss} (bottom) shows the evolution of mole fractions: \ce{CH4} increases while \ce{H2} and \ce{CO2} decrease, particularly in the case  of high  liquid content.

\begin{figure}
\centering
\includegraphics[width=0.485\linewidth]{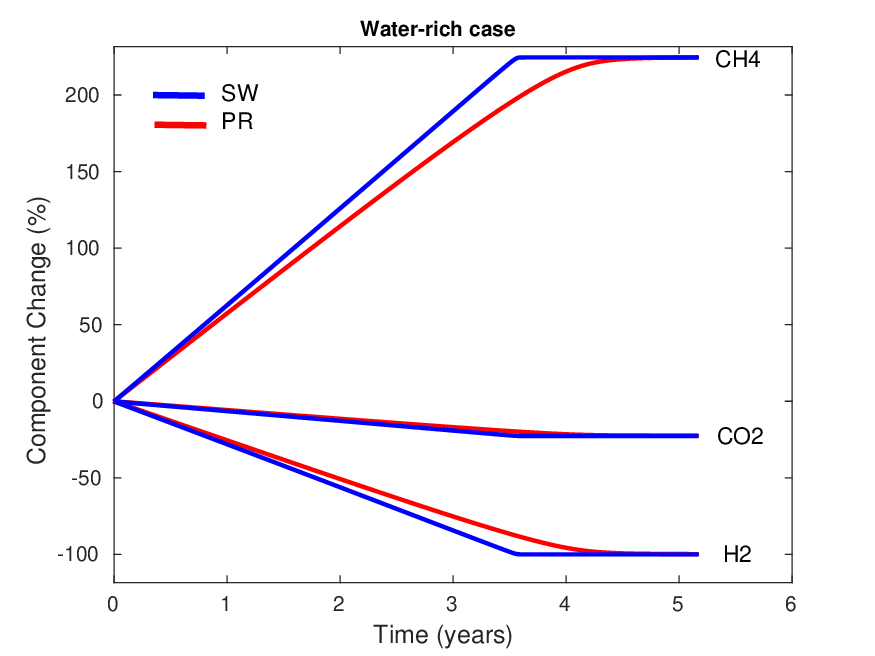}
\caption{1D case:  Comparing PR and SW.}
\label{fig:1d_mole_fractions_PRvsSW}
\end{figure}

In the last experiment, we use compare SW and PR EoSs. The results in the gas-rich case, see
Figure~\ref{fig:1d_mole_fractions_PRvsSW}, indicate a slower consumption of \ce{H2} and \ce{CO2} when using PR EoS,
leading to lower \ce{CH4} production. This result is expected, as the PR is shown to inaccurately capture dissolution
effects, which reduces the extent of gas dissolution into the aqueous phase.

\subsection{Test 2: 2D dome-shaped aquifer}\label{example2D}

\begin{figure}
    \centering
    \includegraphics[width=0.44\textwidth]{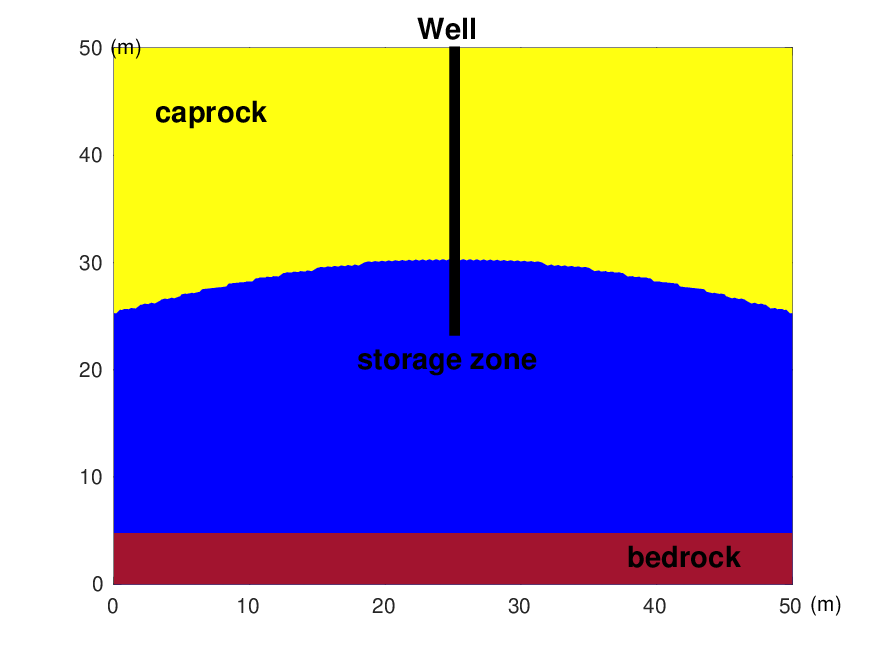}
    \includegraphics[width=0.48\textwidth]{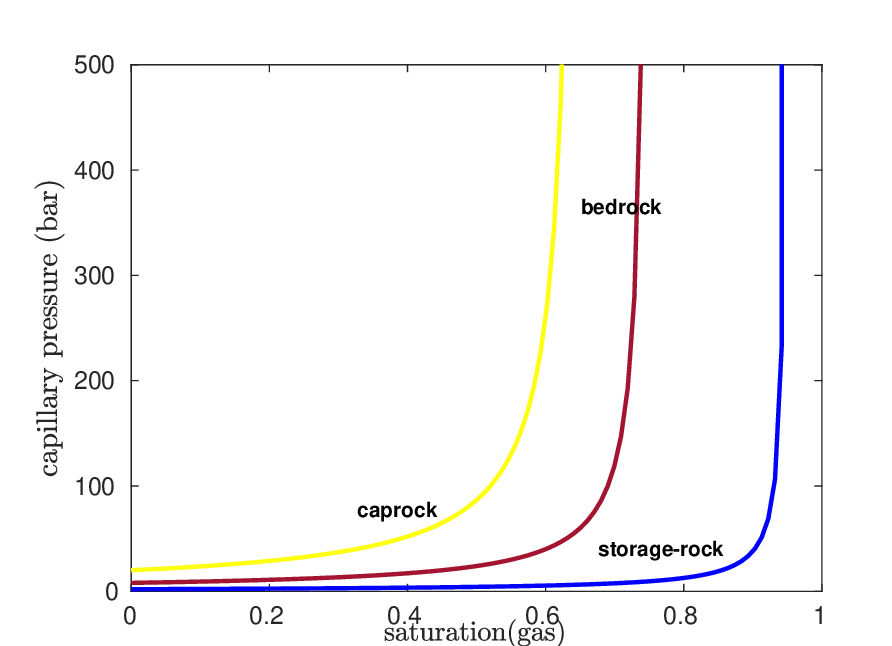}
    \caption{Reproduced from~\cite{AHMED2024104772}: (left) 2D aquifer geometry and (right) capillary pressure curves used in the numerical experiments.}
    \label{fig:domain_case1}
\end{figure}

This example, adapted from~\cite{AHMED2024104772}, investigates hydrogen storage in a saline aquifer using a
compositional model. We analyze the impact of microbial activity on hydrogen plume evolution and loss and \ce{H2}
recoverability. The 2D domain, see Figure~\ref{fig:domain_case1}, consists of a dome-shaped aquifer with caprock and
bedrock layers, extending 50 meters horizontally and in depth. The upper boundary is defined as
\( F(x) = \sigma + r\sin(\pi x) \) with \(\sigma = 25\) and \(r = 5\), ensuring structural trapping.

\begin{table}[h]
\centering
\scriptsize
\begin{tabular}{l l l l l }
\toprule
\textbf{Symbol} & \textbf{Description} &  \textbf{Caprock} & \textbf{Storage zone} & \textbf{Bedrock} \\
\midrule
$\vK$ ($\unit{\milli\darcy}$)  & Permeability  & $1.0\times 10^{-4}$ & 10 & $1.0\times 10^{-2}$ \\
$\Phiref$ (-) & Porosity    & 0.25  & 0.1  & 0.1 \\
$S_{\textnormal{g,r}}$ (-)& Gas residual saturation  & 0.1  & 0.05 & 0.1 \\
$S_{\textnormal{l,r}}$(-) & Water residual saturation  & 0.1  & 0.1  & 0.1 \\
$P_{\textnormal{e}}$ (bar) & Entry pressure  & 10 & 0.4 & 5 \\
$n_l$ (-) & Water exponent   & 4  & 4  & 4 \\
$n_g$ (-) & Gas exponent & 3.5  & 3.5  & 3.5 \\
$\lambda$ (-)& Pore size index   & 1.2  & 1.2  & 1.2 \\
\bottomrule
\end{tabular}
\caption{2D aquifer:  Formation and petrophysical properties~\cite{AHMED2024104772}.}
\label{tab:table_params_illuscase}
\end{table}

Hydrogen retention is ensured by high entry capillary pressure, see Figure~\ref{fig:domain_case1}, and low permeability
in caprock and bedrock. Lateral boundaries impose hydrostatic pressure with a reference pressure \(p_{r} = 40\) bar at
\(z=0\), using \(p_w = p_{r} + \rho g \Delta z\), while top and bottom are no-flux. The domain is initially filled
with brine in hydrostatic equilibrium. Fluid properties are listed in Table~\ref{tab:table_params_illuscase}, with
Brooks-Corey relative permeability and capillary pressure models. We simulate \ch{H2} injection into pure water. The
initial composition is set to 84.8\% water and 15.2\% methane, resulting in an equilibrium saturation of
\([0.1522, 0.8478]\).

\begin{figure}
  \centering
  \includegraphics[width=0.3\linewidth]{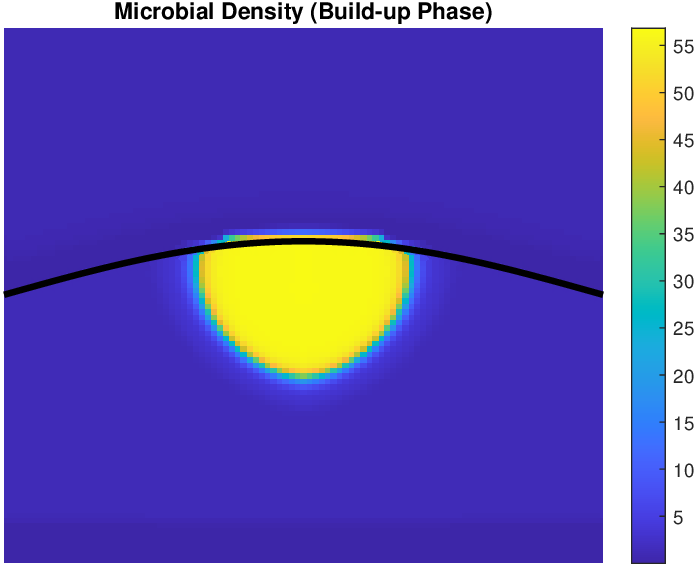}
  \includegraphics[width=0.3\linewidth]{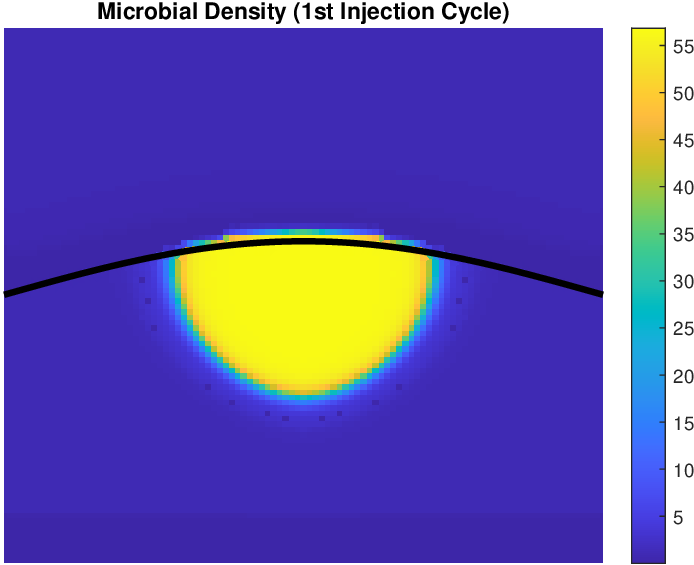}
  \includegraphics[width=0.3\linewidth]{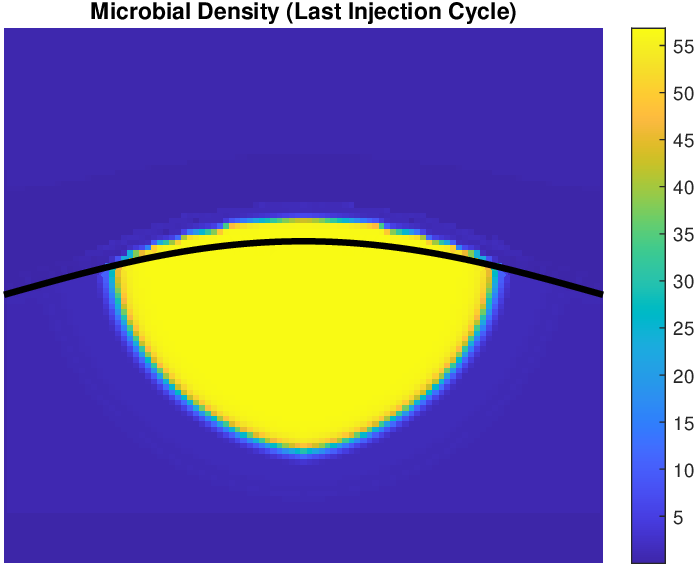}
  \includegraphics[width=0.3\linewidth]{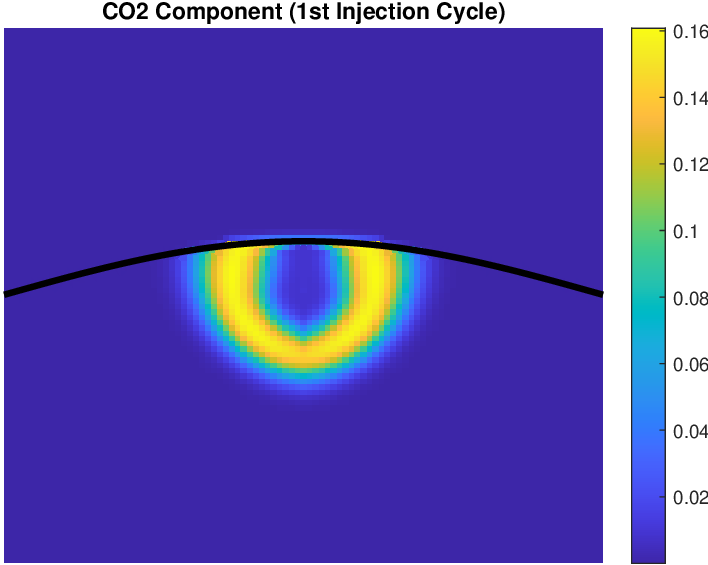}
  \includegraphics[width=0.3\linewidth]{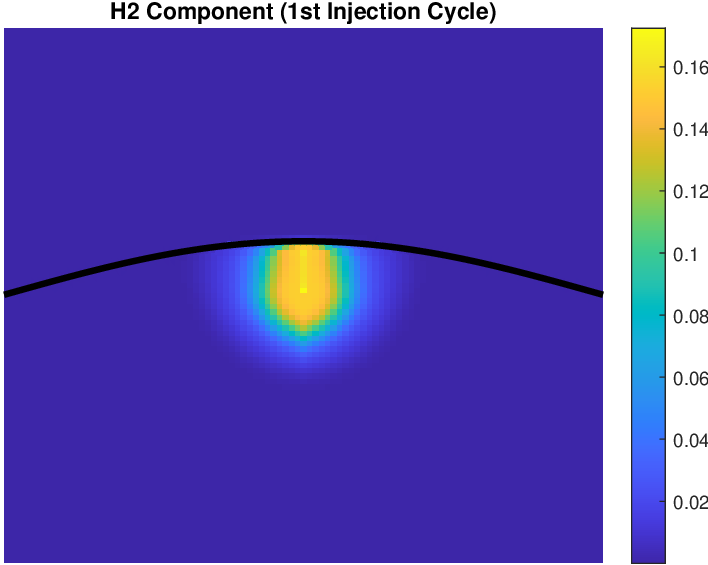}
  \includegraphics[width=0.3\linewidth]{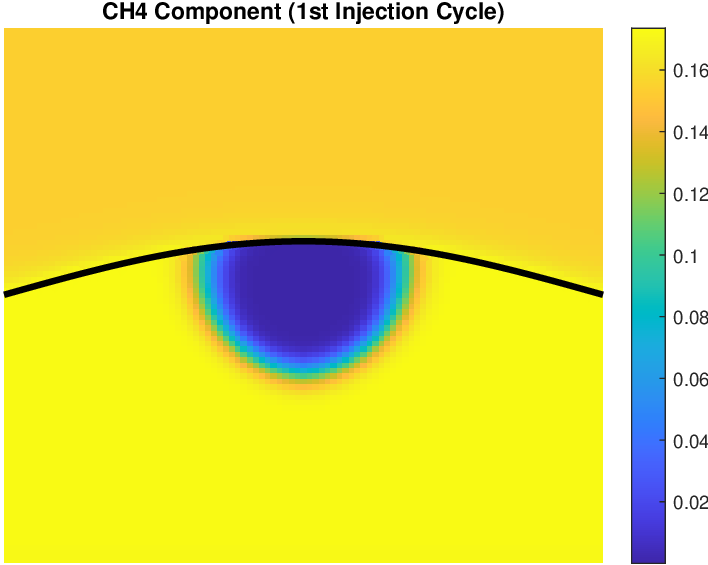}
  \includegraphics[width=0.3\linewidth]{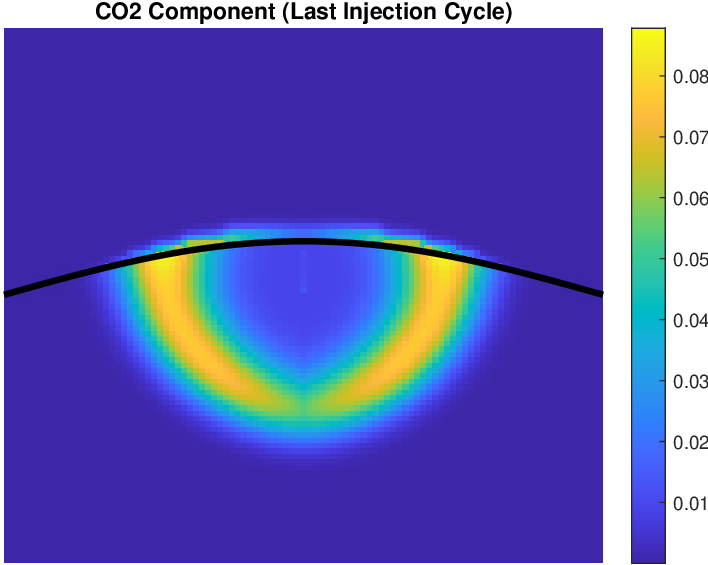}
  \includegraphics[width=0.3\linewidth]{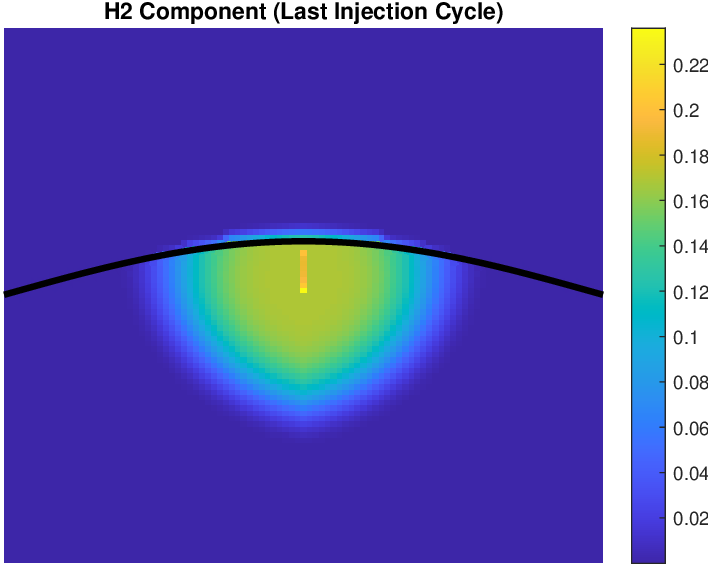}
  \includegraphics[width=0.3\linewidth]{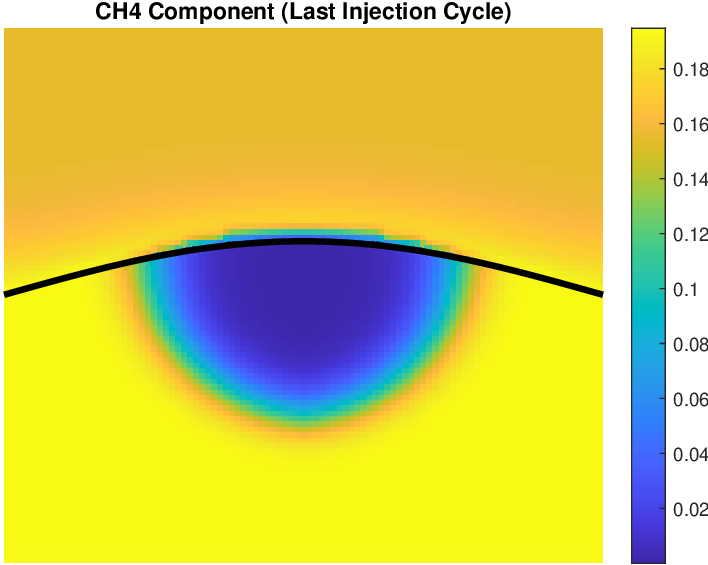}
  \caption{2D aquifer: Distribution of microbial (normalized)  density and mole fractions after the build-up phase, as well as the first and last injection cycles.}
  \label{fig:2d-distribution}
\end{figure}

A single well at the aquifer center has perforations below the caprock. Initially, 90\% of \ce{CO2} and 10\% of \ce{H2}
is injected at \qty{10}{\kilogram\per\day} for 120 days, followed by a 30-day shut-in. A 10-cycle storage sequence (24
months) follows: 30-day pure \ch{H2} injection at \qty{18}{\kilogram\per\day}, a 10-day idle phase, and a 30-day withdrawal at the same rate. Overall, this example is challenging in its physical context and was previously used in~\cite{AHMED2024104772} to study the influence of molecular diffusion on the hydrogen plume, reservoir pressure, and caprock tightness. Here, we investigate the influence of chemical transformation on the \ce{H2} loss and recoverability.

In Figure ~\ref{fig:2d-distribution}, we present the distribution of microbial density (after the build-up phase, first, and last injection cycles) and the corresponding mole fractions of \ce{CO2}, \ce{H2}, and \ce{CH4} (after the first and last injection cycles). Note that the bio-clogging effects are not included here. A distinct gas plume forms after the build-up phase, where the coexistence of \ce{H2} and \ce{CO2} supports microbial activity, which expands over time. During the first injection cycle, with a reduced \ce{CO2} rate, a noticeable decrease
in \ce{CO2} occurs near the wellbore (maximum \ce{CO2} mole fraction is halved). A highly mixed gas saturation is
observed from the component plumes. By the final injection cycle, the maximum mole fraction of \ce{CH4} increases from
0.16 to 0.2, indicating enhanced microbial conversion. Additionally, dissolved \ce{H2} infiltrates the caprock, but its
amount is reduced due to microbial activity (see the top-left figure for microbial activity).

\begin{figure}
    \centering
    \includegraphics[width=0.32\linewidth]{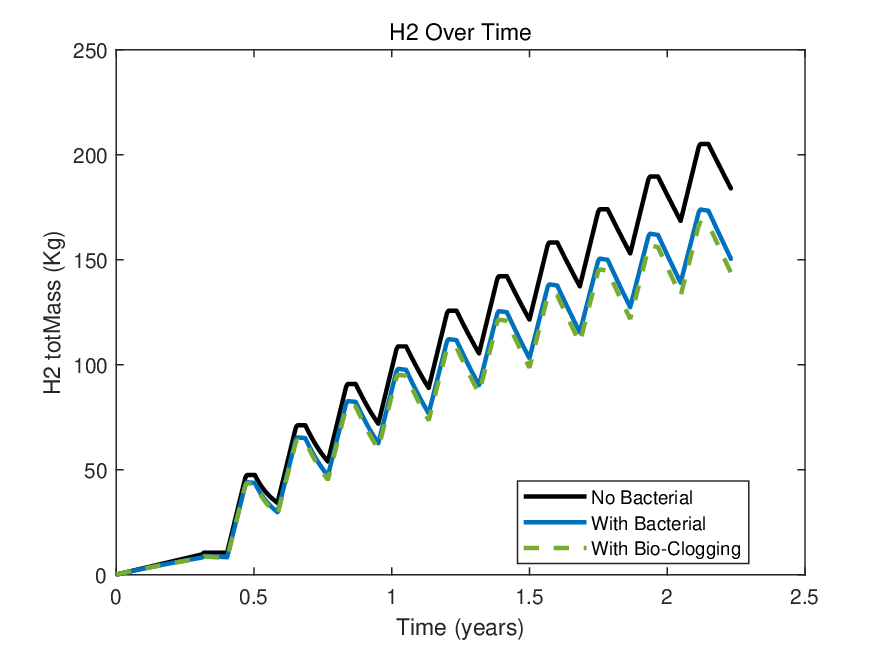}
    \includegraphics[width=0.32\linewidth]{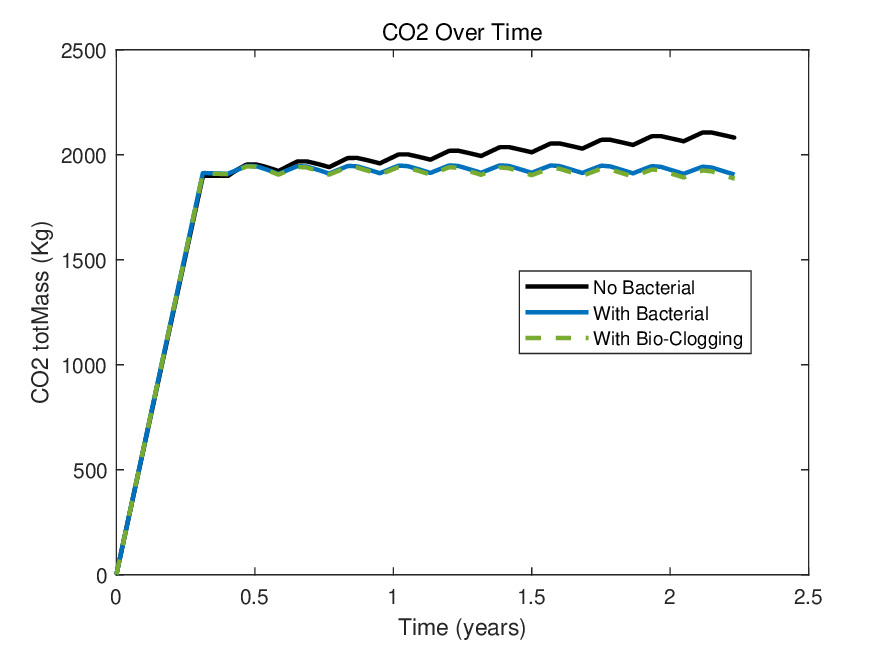}
    \includegraphics[width=0.32\linewidth]{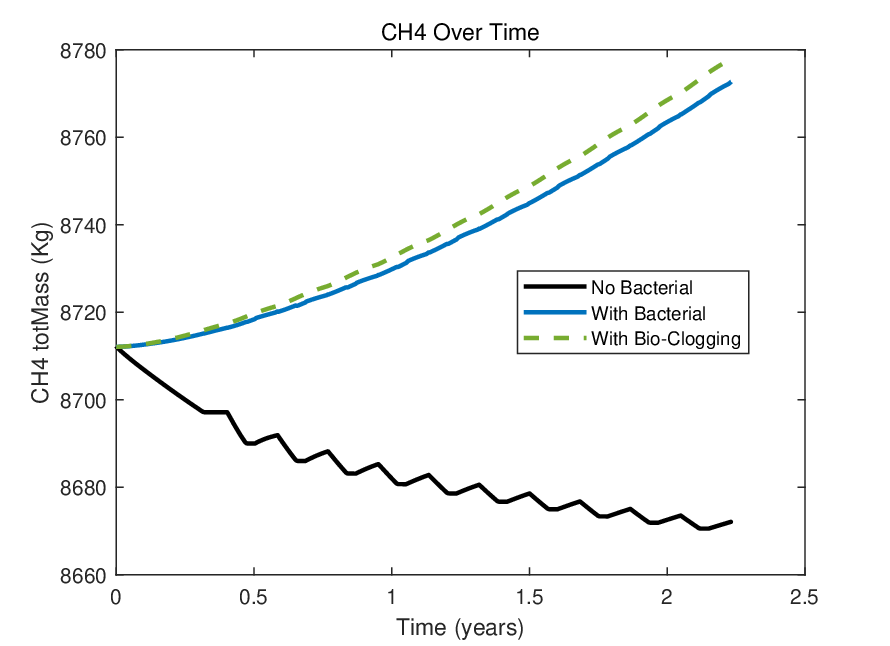}
        \includegraphics[width=0.32\linewidth]{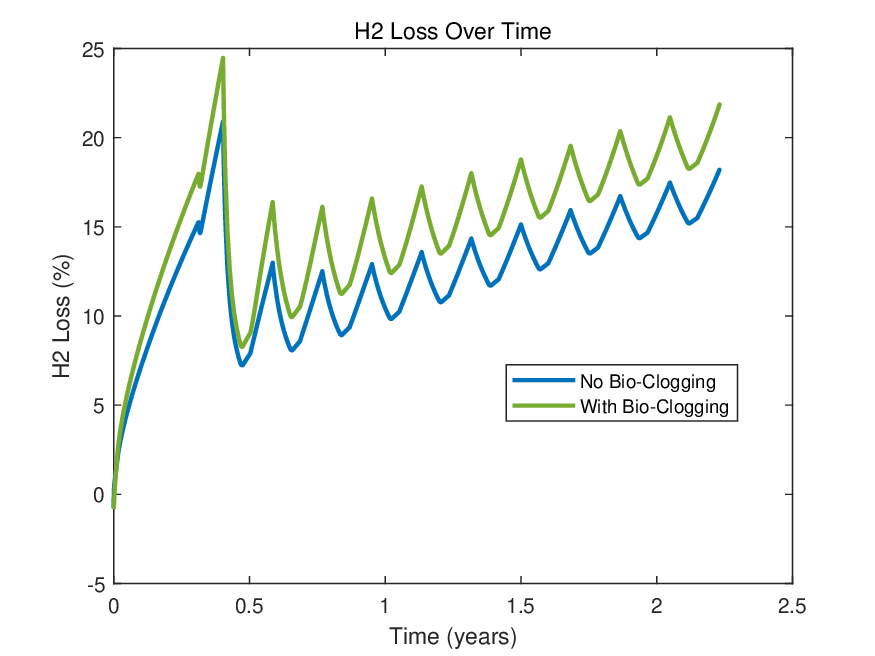}
    \includegraphics[width=0.32\linewidth]{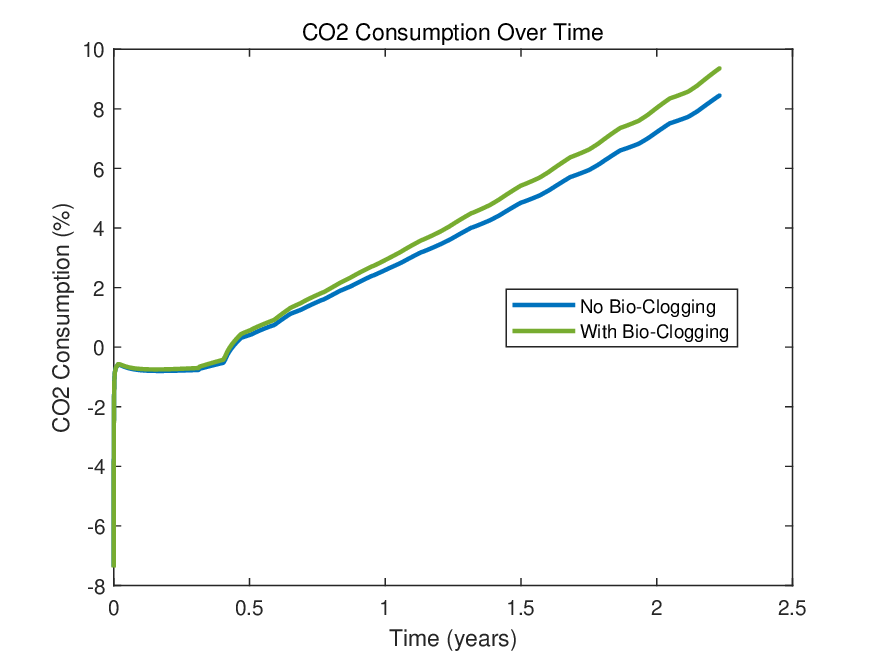}
    \includegraphics[width=0.32\linewidth]{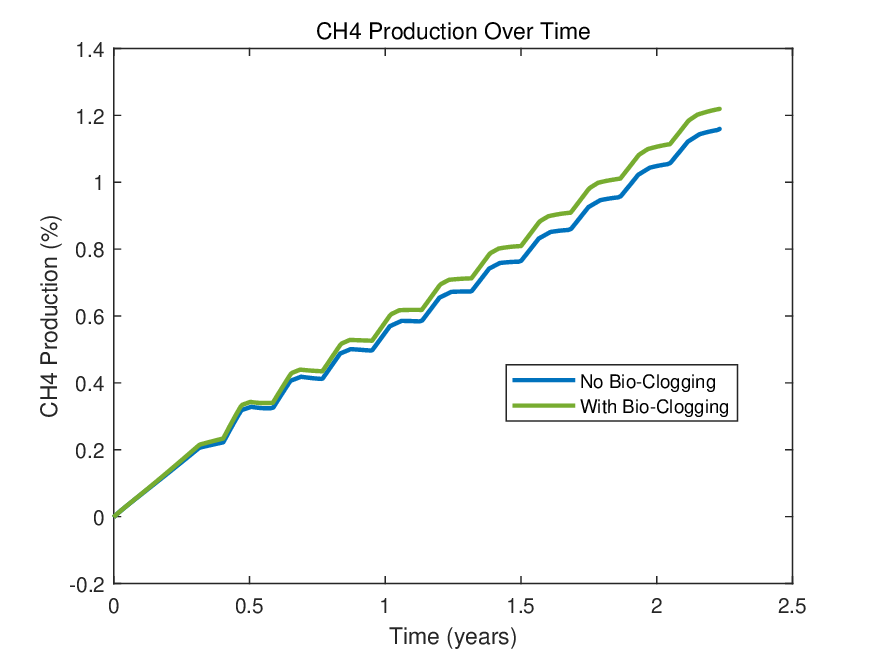}
            \includegraphics[width=0.32\linewidth]{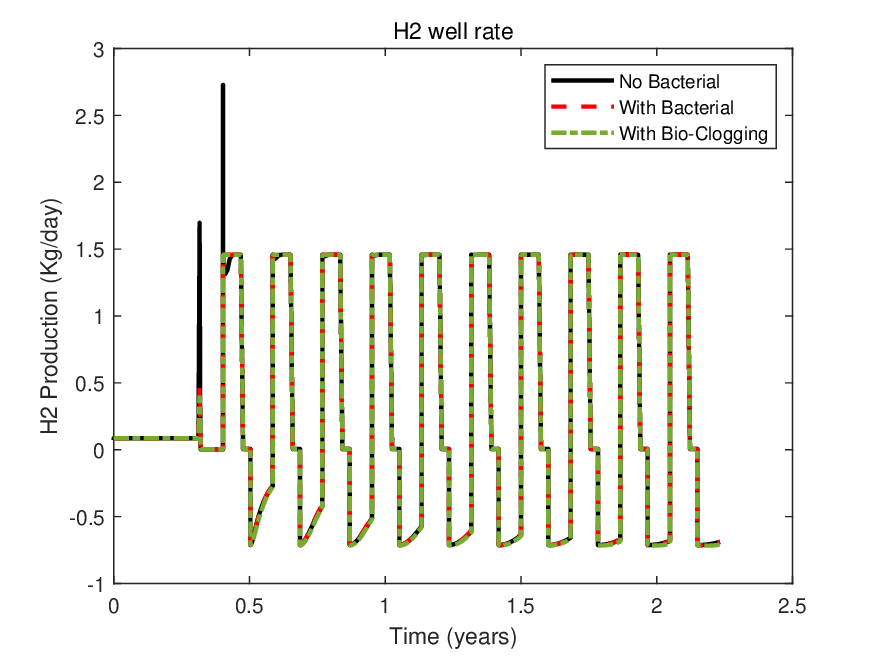}
    \includegraphics[width=0.32\linewidth]{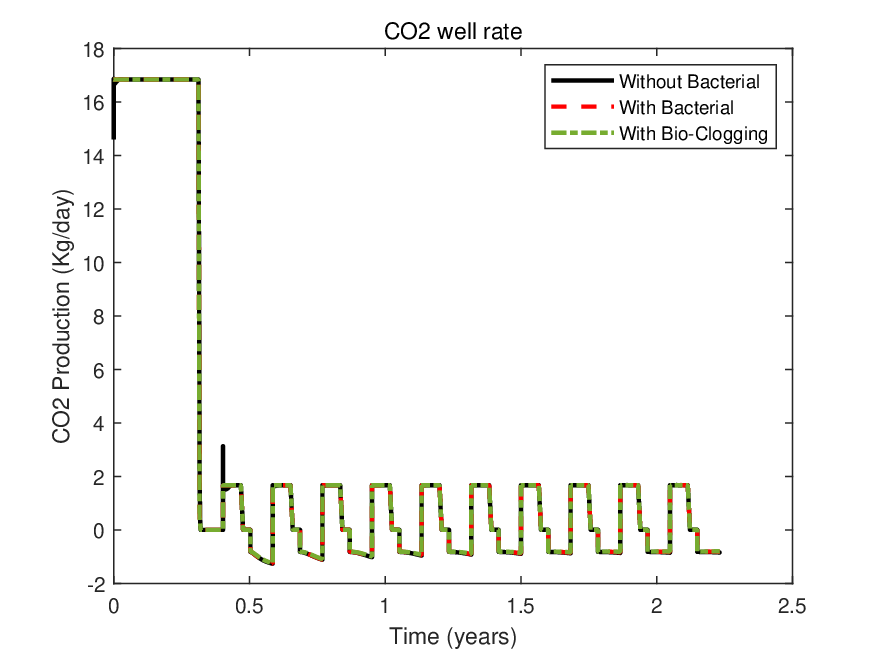}
    \includegraphics[width=0.32\linewidth]{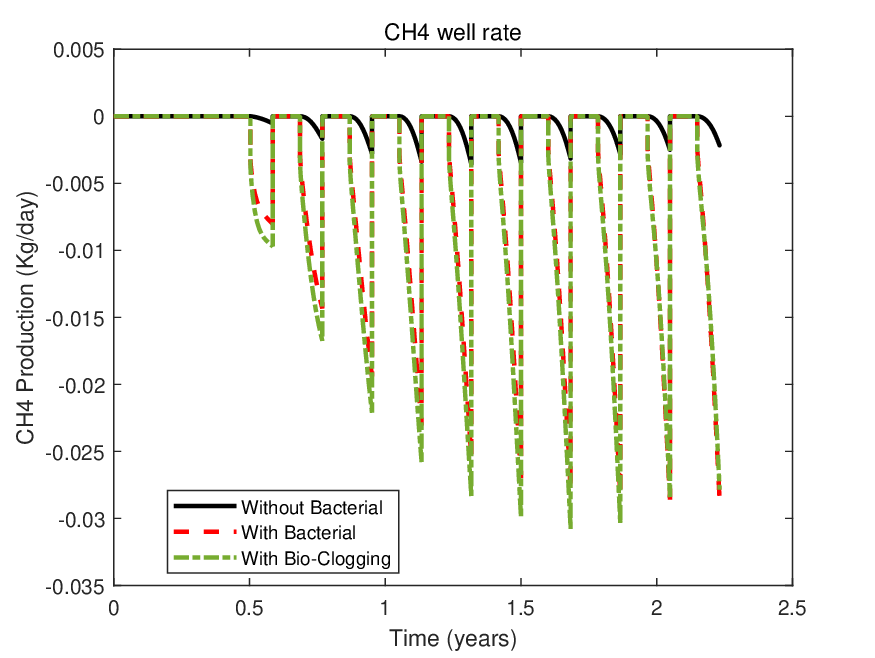}
    \caption{2D aquifer:  Impact of biochemical transformation on cyclic \ce{H2} injection.}
    \label{fig_2d:Impact_biochemical_transformation}
\end{figure}

We present in Figure~\ref{fig_2d:Impact_biochemical_transformation} the impact of microbial activity on the variation in
component total mass (top), chemical loss and production (middle), and well efficiency (bottom). Methanogenesis occurs
as soon as the simulation starts, resulting in an \ce{H2} loss of 18\% by the end of the simulation. While the total
\ce{H2} loss does not significantly impact production efficiency, the purity of the produced \ce{H2} is affected by the
increased \ce{CH4} production. However, bio-clogging increases chemical transformation by extending residence time,
which promotes more reactions and reduces \ce{H2} migration. Table~\ref{tab_2d:h2_co2_ch4} presents the variation in the
mixture at the end of the simulation.

In all cases, \ce{H2} loss due to solubility remains around 1.5\% and may decrease with dissolved salts. Microbial
activity induces similar transformations even at high salinity, suggesting control by yield, decay, and gas
composition. Pressure and temperature, within favorable ranges for methanogens, influence dissolution like salinity but
have a minor effect on \ce{H2} loss, consistent with~\cite{SAFARI2024120426}.

\begin{table}[h]
\centering
\scriptsize
\begin{tabular}{l l l l}
\toprule
& \textbf{\ce{H2} loss (\%)} & \textbf{\ce{CO2} consumption (\%)} & \textbf{CH$_4$ production (\%)} \\ 
\midrule
\textbf{No clogging} & 18.21 & 8.44 & 1.20 \\ 
\textbf{Bio clogging} & 21.68 & 9.36 & 1.22 \\ 
\bottomrule
\end{tabular}
\caption{2D aquifer: Comparison of \ce{H2} loss, \ce{CO2} consumption, and CH$_4$ production.}
\label{tab_2d:h2_co2_ch4}
\end{table}

\begin{figure}
    \centering
    \includegraphics[width=0.32\linewidth]{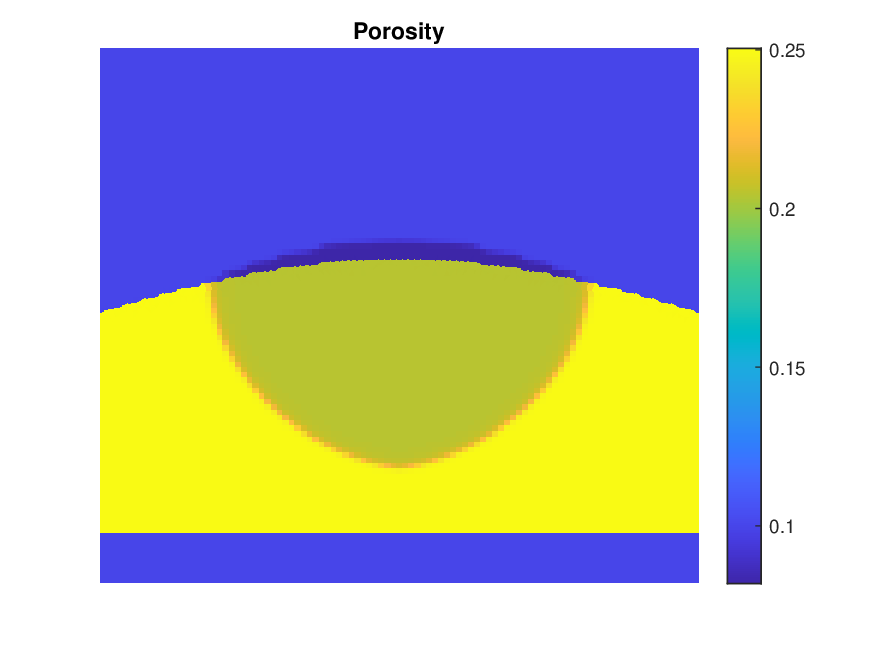}
    \includegraphics[width=0.32\linewidth]{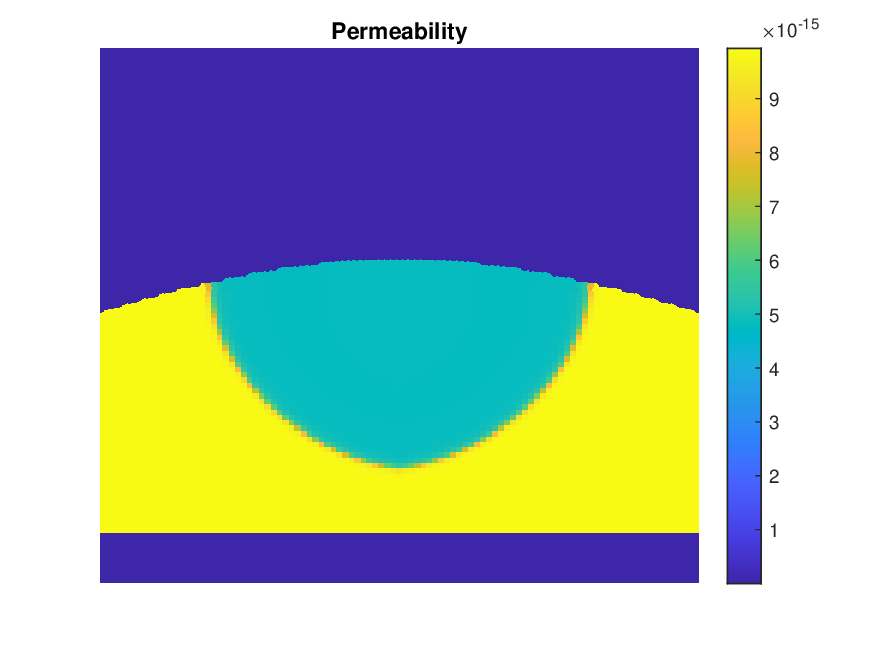}
    \caption{2D aquifer:  bio-clogging due to biochemical transformation.}
    \label{fig_2d:bioclogging}
\end{figure}

\subsection{Test 3: underground hydrogen storage in a 3D regular reservoir}\label{Benchmark2023}
This studied case is based on the geometry and scenario of a test initially proposed by \cite{Khoshnevis2023}. It consists in studying the transport and evolution of components (\ce{H2}, \ce{CO2}, \ce{CH4}, \ce{H2O}) and micro-organisms of a two-phase system in a reservoir 3D in which \ce{H2} and \ce{CO2} (cushion gas) are injected. In the present work, the reservoir is considered as an aquifer containing either pure water or salt water.

\subsubsection{Simulation setup}
The top of the reservoir is located at $1000\unit{\meter}$ deep and its horizontal dimensions are $1525\unit{\meter}\times1525\unit{\meter}$ while its vertical dimension is $50\unit{\meter}$. Then, it is composed of an homogeneous porosity $\Phi=0.2$ and an heterogeneous absolute permeability $\mathbf{K}=[100\unit{\milli\darcy},100\unit{\milli\darcy},10\unit{\milli\darcy}]$. It is perforated by a vertical well placed on the center of the top face of the reservoir.\\
The system is initialized so as to enable the methanogenic archae activity, i.e.  $P_0=100$ \unit{\bar}, $T_0=313.15\,\unit{\kelvin}$ and $S_{l0}=0.2$. Initially, the reservoir is composed of a liquid phase and a gas phase which contain three components: water $z_{0,\ce{H2O}}=0.7$, carbon dioxide $z_{0,\ce{CO2}}=0.02$ and methane $z_{0,\ce{CH4}}=0.28$.

Furthermore, the fluid properties (see table \ref{tab:fluidproperties}) are those of water for the liquid phase and those of air at the surface atmosphere conditions, namely $P_{\text{atm}}=10^5\, \unit{\pascal}$ and $T_{atm}=283.15\,\unit{\kelvin}$. It is to be noted that the air has been approximated by a gas composed of $80\%$ \ce{N2} and $20\%$ \ce{O2} and that the data come from the \href{https://webbook.nist.gov/chemistry/fluid/}{NIST webbook}. 
\begin{table}[h!]
\centering
\scriptsize 
\caption{Fluid properties in each phase.}
\label{tab:fluidproperties}
\begin{tabular}{l l l l }
\toprule
Phase & Density ($\mathrm{kg} .\mathrm{m}^{-3}$) & Viscosity ($\mathrm{Pa} .\mathrm{s}$) & compressibility ($\mathrm{Pa}^{-1}$)\\
\midrule
liquid & $999.7$ & $1.3059 \times 10^{-3}$ & $5.0015 \times 10^{-10}$  \\
gas & $1.2243$ & $1.763 \times 10^{-5}$ & $1.0009 \times 10^{-5}$  \\
\bottomrule
\end{tabular}
\end{table}

Relative permeabilities are determined using the Brooks-Corey model~\cite{BrooksCorey1964}, with exponent parameters $(\lambda_l, \lambda_g) = (2,2)$, entry pressure $P_{\textnormal{en}} = 10^4\,\unit{\pascal}$ and residual saturations $(S_{lr}, S_{gr}) = (0.2, 0.05)$ for each phase. The microbial diffusion effect is neglected here.

Homogeneous Neumann boundary conditions are imposed. A cartesian mesh is used for the reservoir with $31\times31\times8$ cells, corresponding to each spatial direction. For solving the linearized systems of equations that occur, the default LU factorization or the ILU(0) preconditioned GMRES~\cite{mrst-book-2-ad} method proposed by MRST are used. Furthermore, the computational performance is improved thanks to the accelerated AD backend of MRST. A fixed time step of $1\unit{\day}$ is chosen to perform the simulations.

Firstly, a gas mixture composed of $60\%$ \ce{H2} and $40\%$ \ce{CO2} is injected through the well at a rate of $10^6 \unit{\meter^3\per\day}$ for a period of build-up of $60$ days). This is followed by a rest period of $20$ days and a second gas mixture injection period of $30$ days, composed of $95\%$ \ce{H2} and $5\%$ \ce{CO2}. Then comes a period of idle $20$ days, a production period of $30$ days at a rate of $10^6\,\unit{\meter^3\per\day}$, and a final period of idle $20$ days. This scenario close to the one presented in \cite{Khoshnevis2023} is extended to a scenario of 6 cycles to analyze the long-term impacts. For the first cycle the same parameters are kept meanwhile in the following cycles there are neither a build-up period nor a rest period.

Several simulation cases are conducted to estimate the effect of methanogenic archae activity on the multi-component liquid-gas system during the 6 cycles. In the first case, a pure water aquifer without microbial activity is considered, while in the second case, the system is initialized with a microbial population of $\bactN_0=1$ (initial density $n_0=10^{8}\unit{\per\cubic\meter}$) in the pure water aquifer. As well, the performances of the linear solvers LU factorization and the ILU(0) preconditioned GMRES are compared.

In the third case, a saltwater aquifer without microbial activity is considered, while in the fourth case, the system involves microbial activity in a saltwater aquifer. In the fifth case (resp. sixth case), the initial microbial density is put to $n_0=10^{9}\unit{\per\cubic\meter}$ in a pure water aquifer (resp. salt water aquifer). In the sixth case, the molecular diffusion effects are are taken into account in a pure water aquifer and microbial activity. 

\subsubsection{Results and discussion}
\begin{figure}
    \centering
    \includegraphics[width=0.48\linewidth]{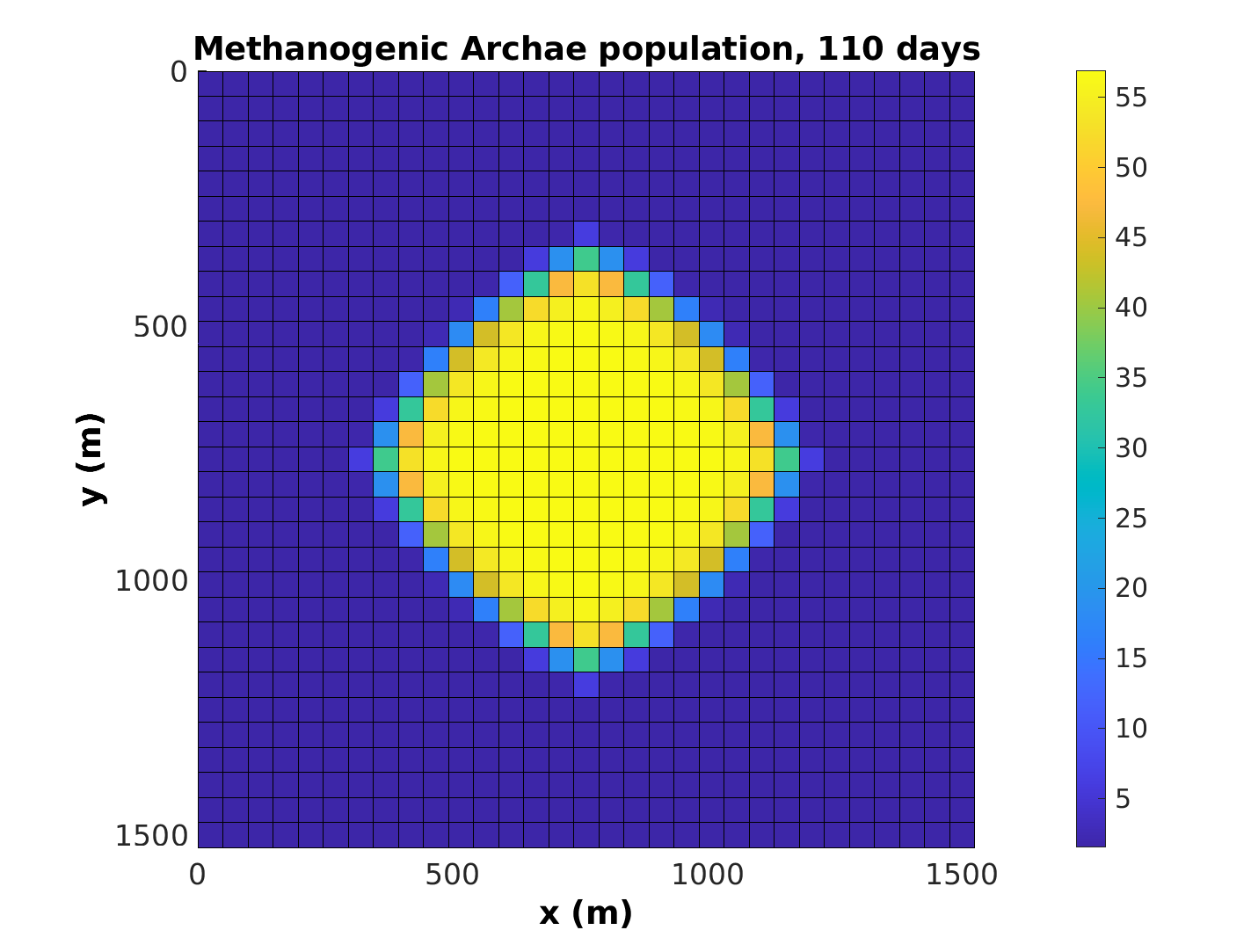}
    \includegraphics[width=0.48\linewidth]{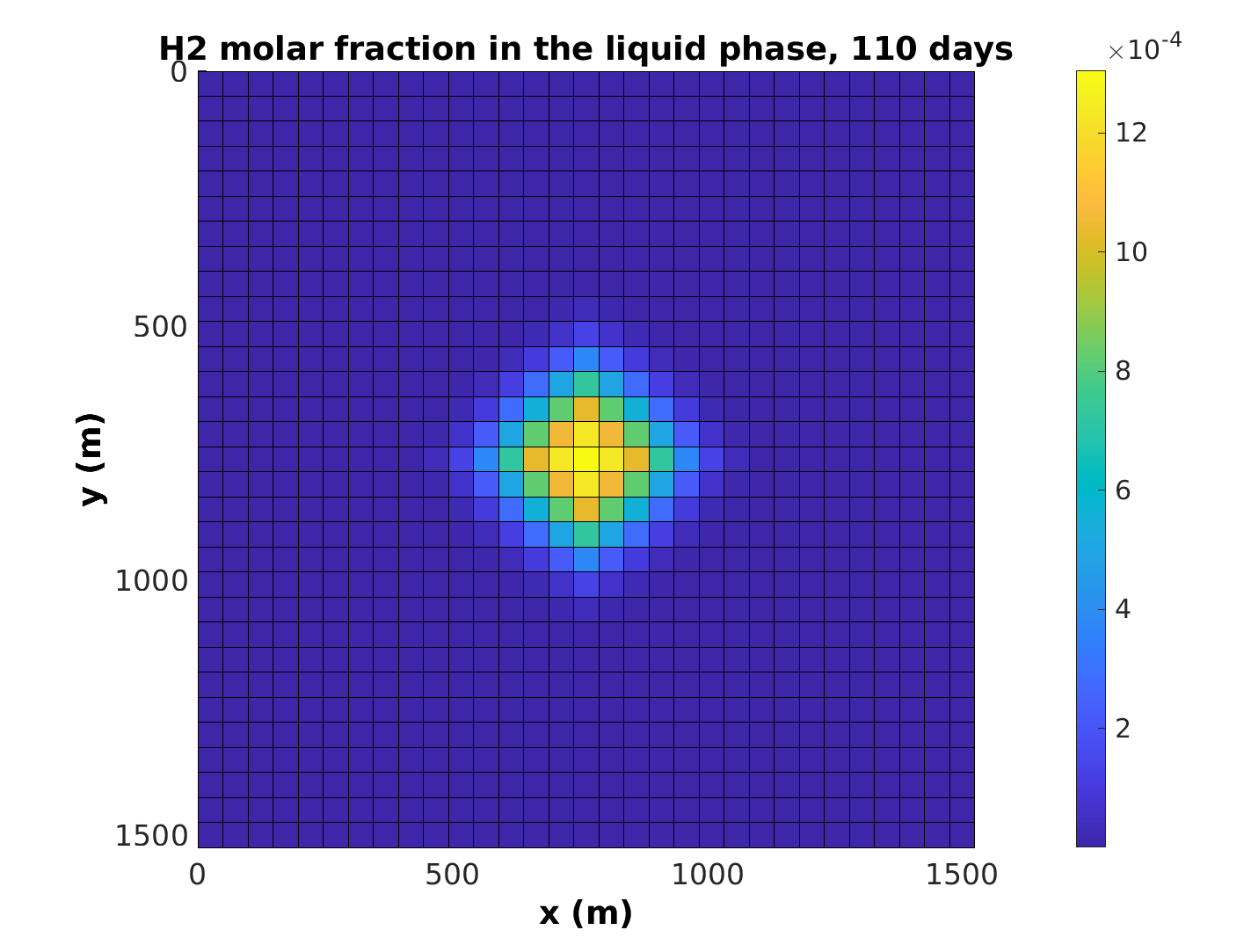}
    \caption{Test 3: microbial activity (left) and \ce{H2} solubility (right) in the vicinity of the well at the end of the first injection period.}
    \label{fig:nbactinjectdays}
\end{figure}

\begin{figure}
   \includegraphics[width=0.48\linewidth]{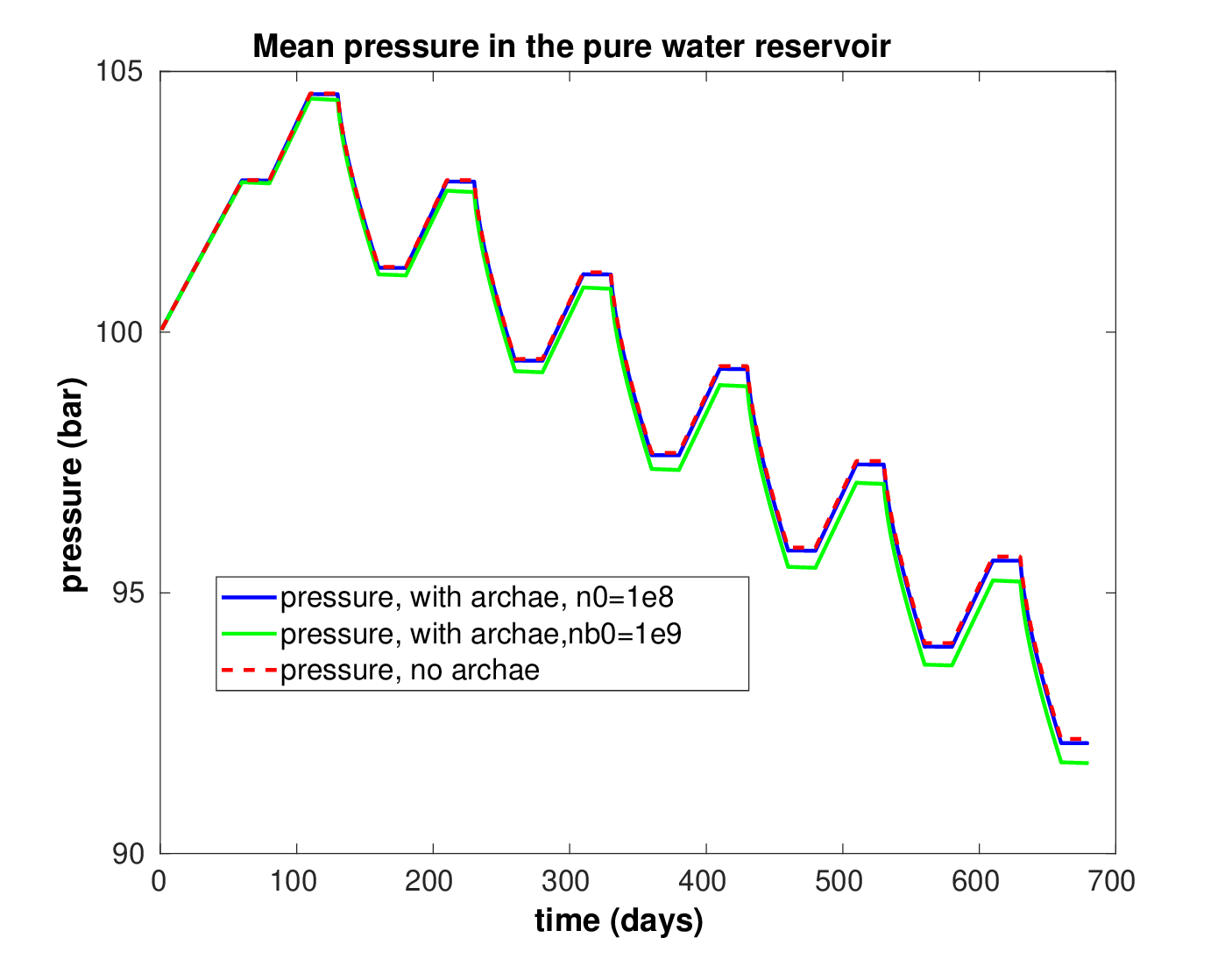}
    \includegraphics[width=0.48\linewidth]{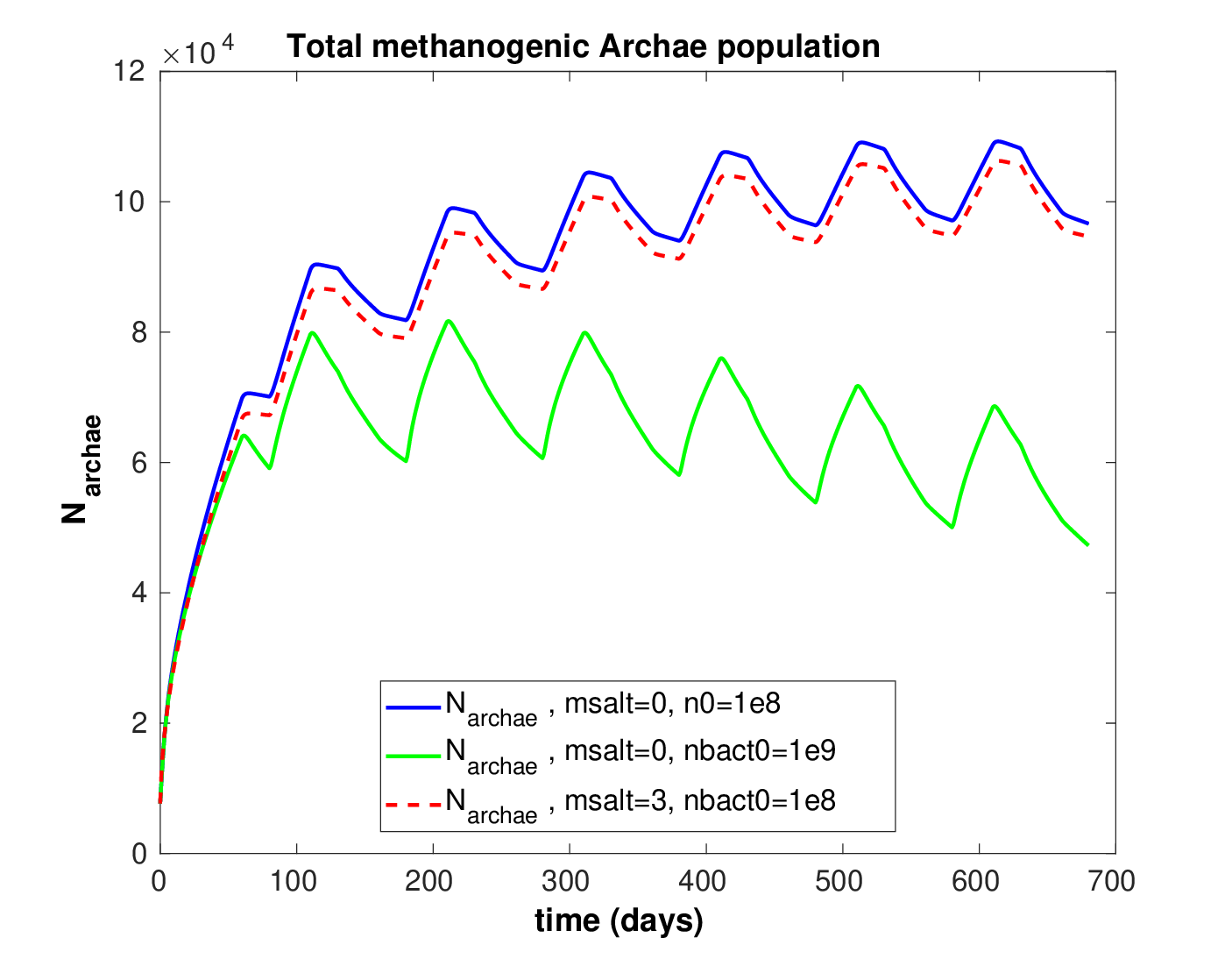}
   \caption{Test 3: evolution of the mean pressure in the reservoir(left). Evolution of total microbial population (right) in a pure water and salt water aquifer (right) during 6 cycles.}
  \label{fig:pressure_nbact}
\end{figure}

Firstly, we aim at studying the impact of the microbial activity on the UHS in a pure water ($\msalt=0$) aquifer. Hence, Figure \ref{fig:nbactinjectdays} shows that microbial activity occurs in the neighborhood of the well rich in \ce{H2} and that the population of microorganisms varies along with the amount of available \ce{H2} and
\ce{CO2}.  The simulation results show a close connection between the activity of methanogenic archae and the amount of \ce{H2} and \ce{CO2} available in the reservoir. In fact, an increase in the gases \ce{H2} and \ce{CO2} during the injection period induces a growth of the microbial population. In the same way, a decay of the microbial population is noted during the production period, see Figure \ref{fig:pressure_nbact} (right). It is also to be noted that the solubility of \ce{H2} (resp \ce{CO2}) in water reaches $1.3\times 10^{-3}$ (resp $9.8\times 10^{-3}$) near the well, which is consistent with the results of section \ref{subsection:SW_EoS} in a pure water aquifer.

\begin{figure}
  \begin{subfigure}{0.33\textwidth}
    \includegraphics[width=1.0\linewidth]{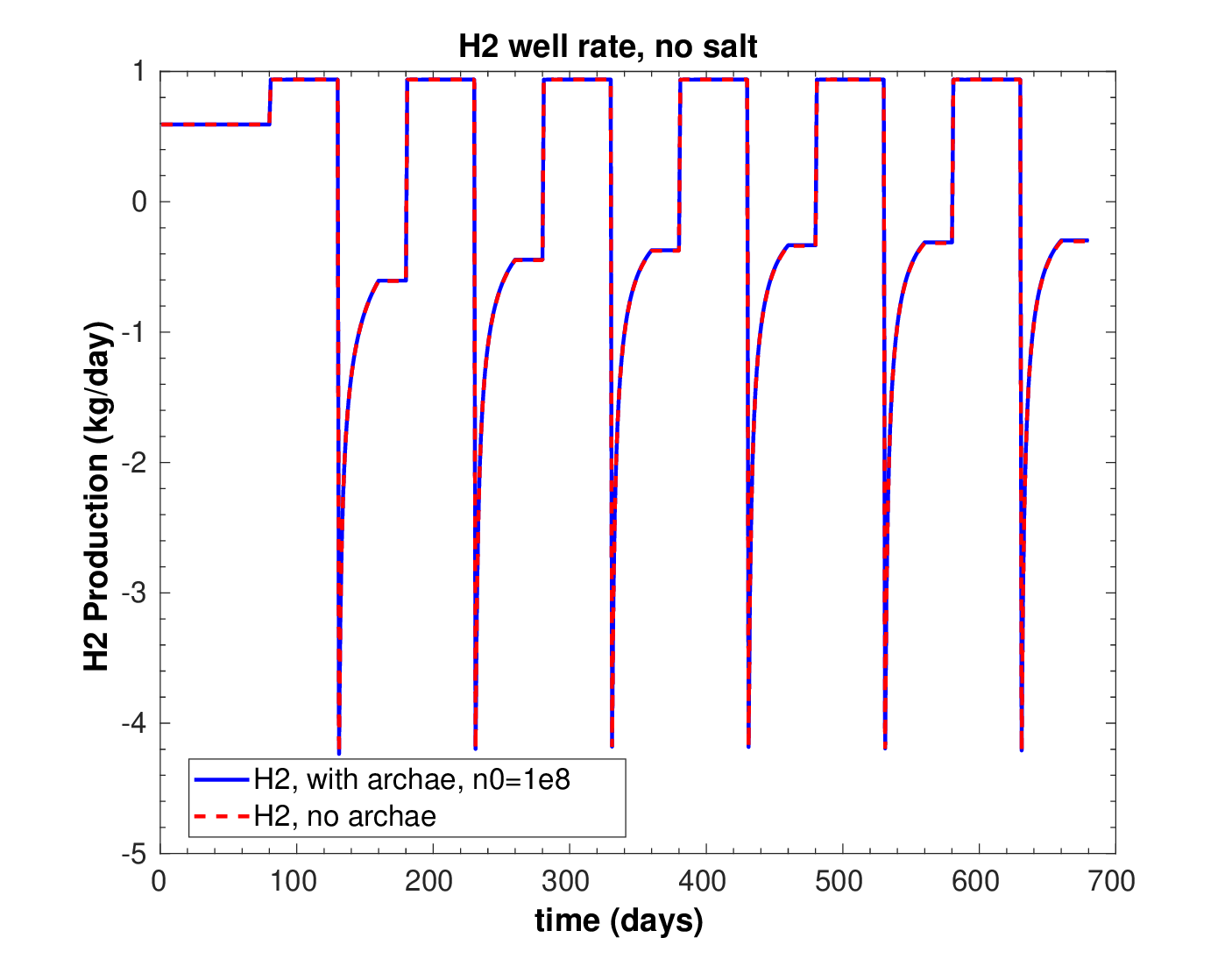}
    \caption{ }
    \label{fig:H2_well_nosalt_6cycles}
  \end{subfigure}\hfill
  \begin{subfigure}{0.33\textwidth}
    \includegraphics[width=1.0\linewidth]{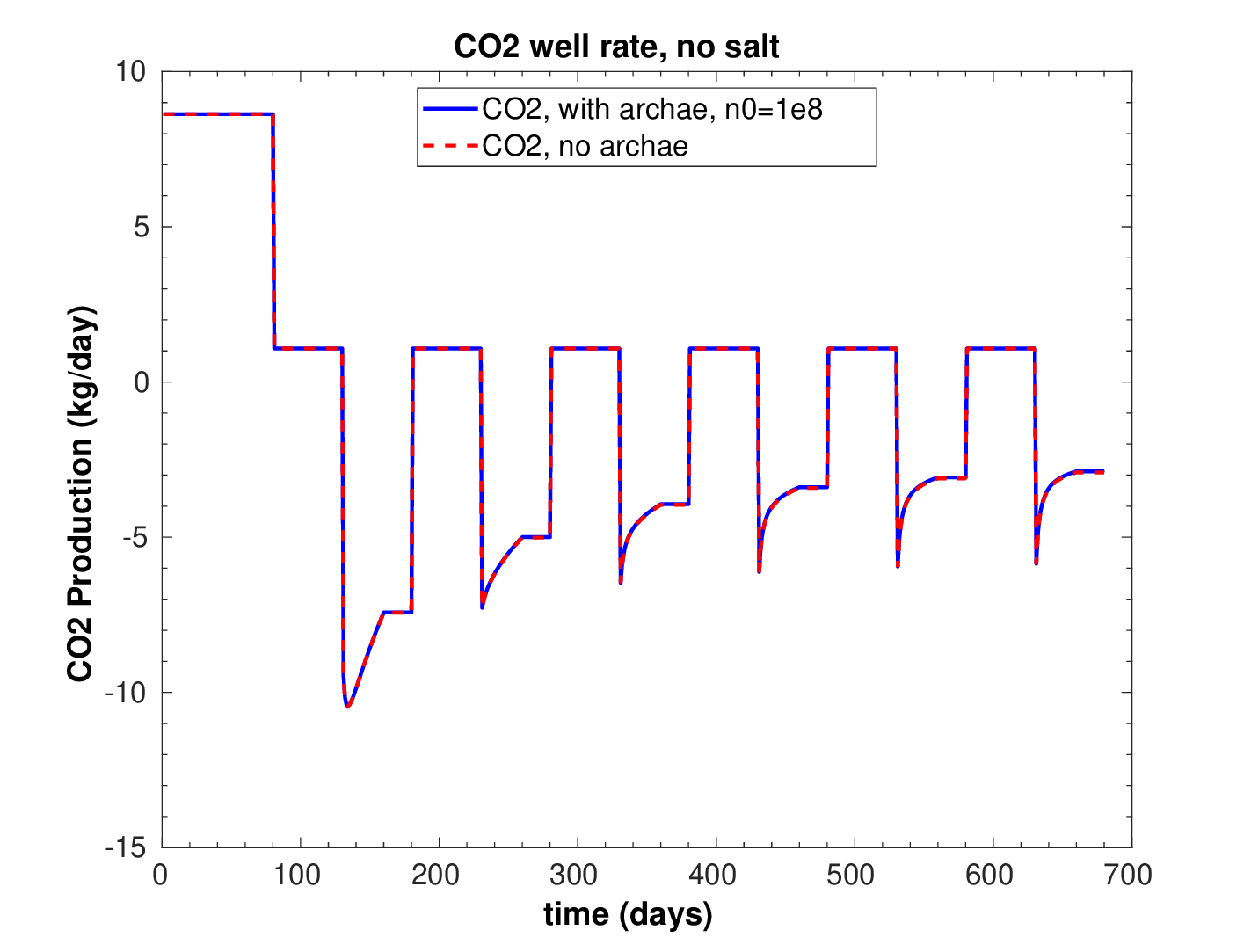}
    \caption{ }
    \label{fig:CO2_well_nosalt_6cycles}
  \end{subfigure}
  \begin{subfigure}{0.33\textwidth}
    \includegraphics[width=1.0\linewidth]{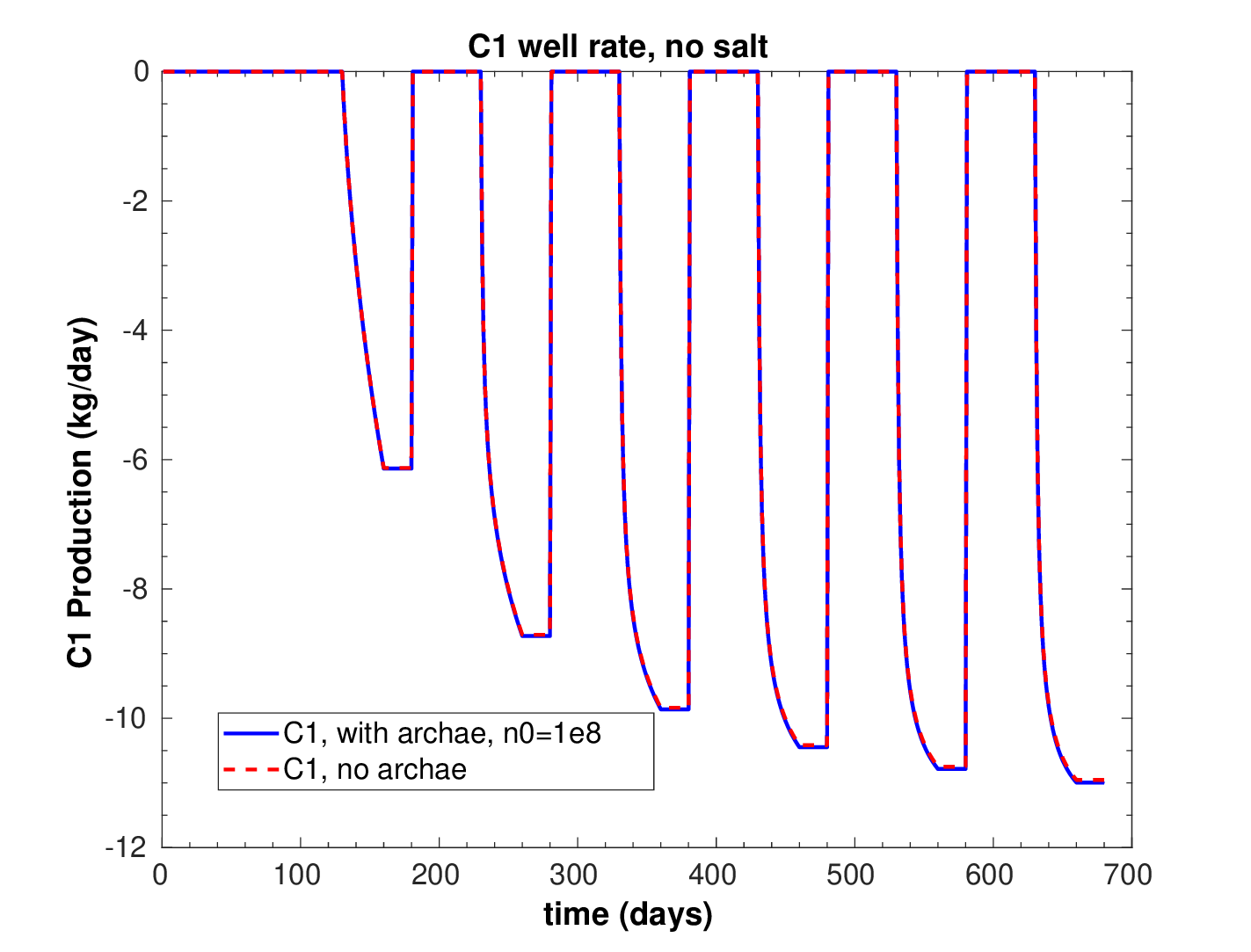}
    \caption{ }
    \label{fig:C1_well_nosalt_6cycles}
  \end{subfigure}

 \caption{Test 3: \ce{H2}, \ce{CO2} and \ce{C1}  production in the well of the pure water aquifer.}
  \label{fig:componentswell_nosalt_6cycles}
\end{figure}

\begin{figure}
\begin{subfigure}{0.33\textwidth}
\includegraphics[width=1.0\linewidth]{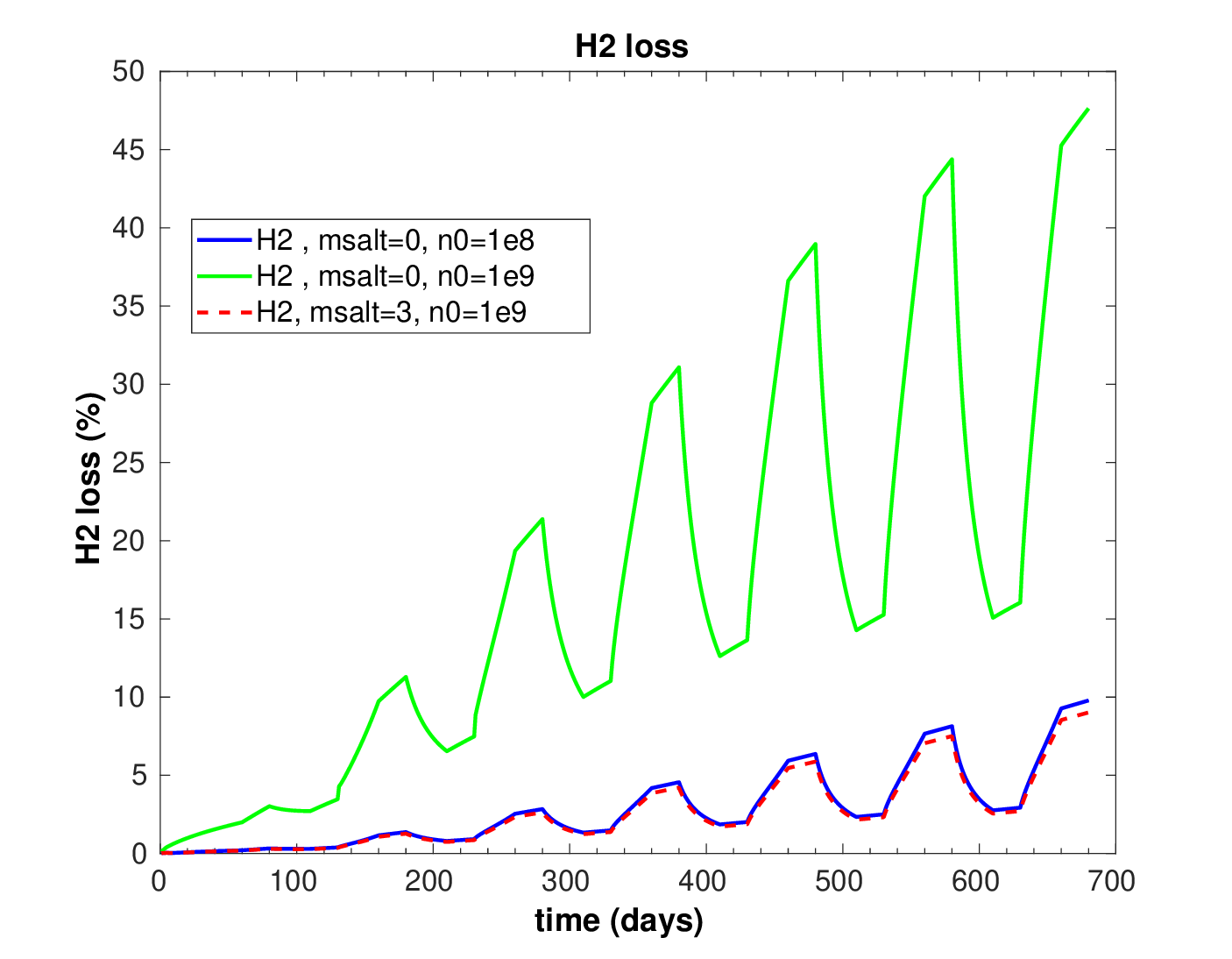}
\caption{ }
\label{fig:H2_loss_comparsalt_6cycles}
\end{subfigure}\hfill
\begin{subfigure}{0.33\textwidth}
\includegraphics[width=1.0\linewidth]{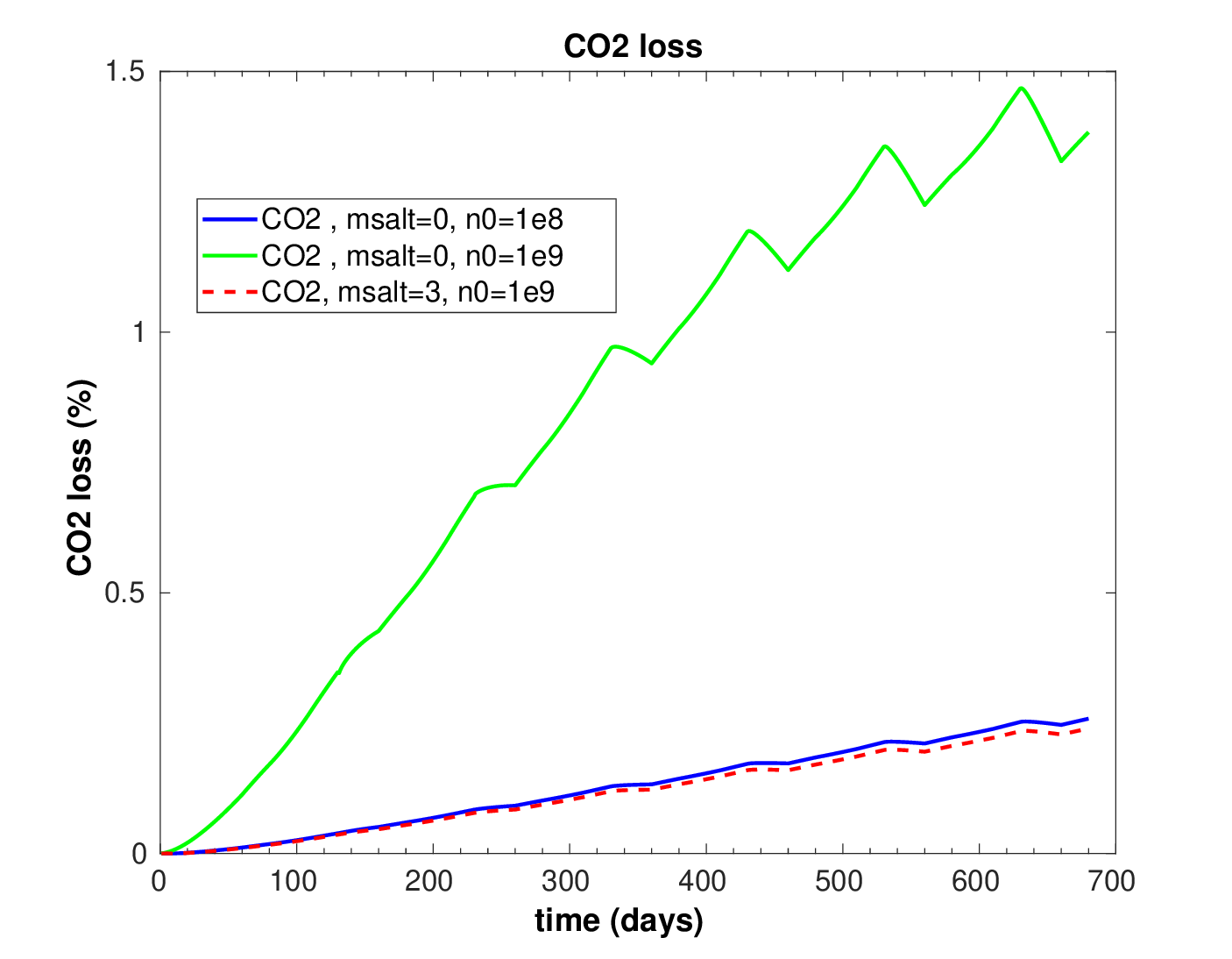}
\caption{ }
\label{fig:CO2_loss_comparsalt_6cycles}
\end{subfigure}
\begin{subfigure}{0.33\textwidth}
\includegraphics[width=1.0\linewidth]{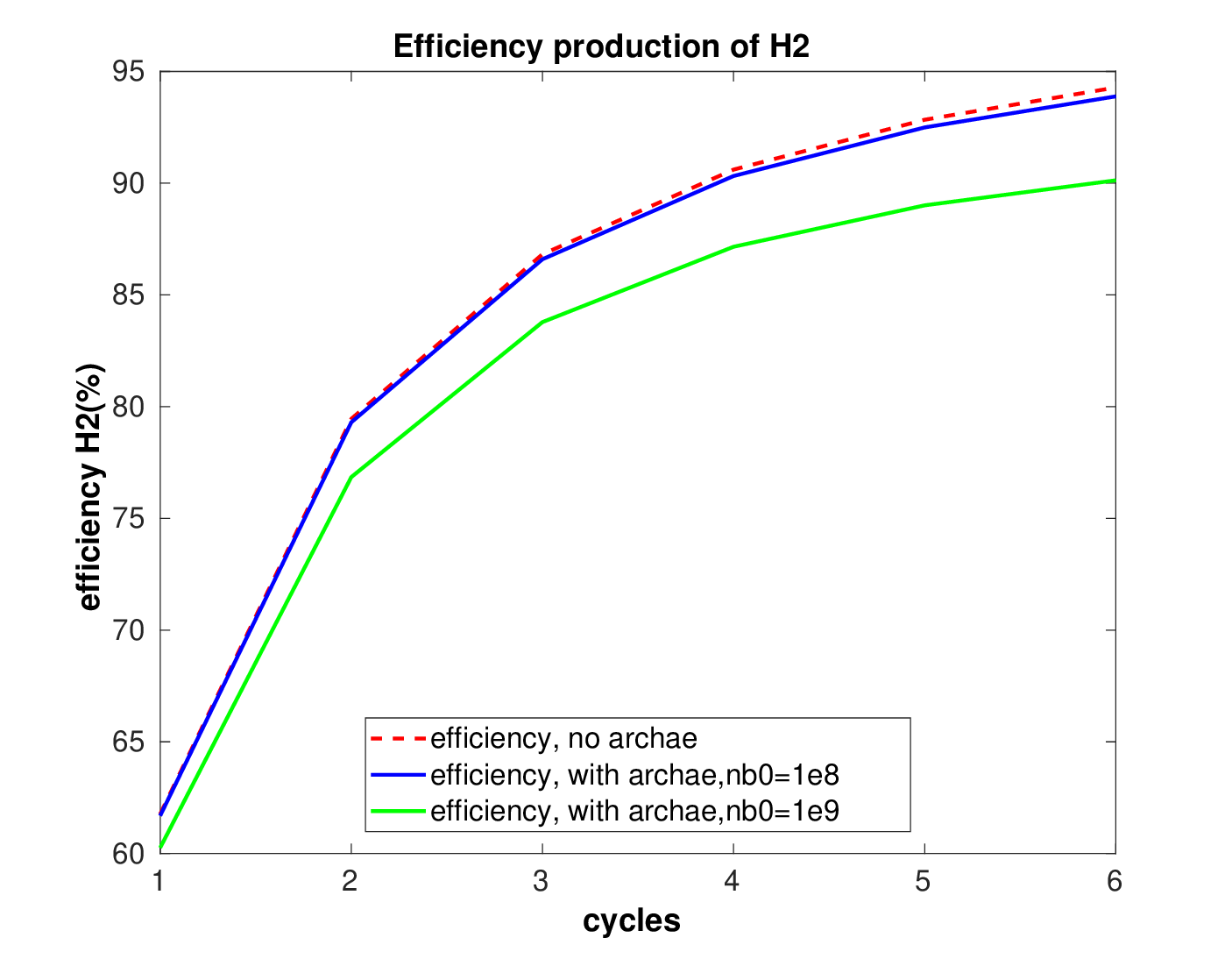}
\caption{ }
\label{fig:efficiencyH2_6cycles}
\end{subfigure}
\caption{\ce{H2} loss, \ce{CO2} consumption and \ce{H2} efficiency production in the well of both pure and salt water aquifer during 6 cycles.}
\label{fig:components_well_lossproduct_6cycles}
\end{figure}

Moreover, Figure \ref{fig:pressure_nbact} shows the impact of microbial activity in the sense that the reaction process induces a decrease in pressure. This decrease can explain a decrease in production efficiency. In fact, the efficiency of production is calculated in both cases as the total mass of \ce{H2} produced during the production period divided by the total mass of \ce{H2} injected into the well (see Figure \ref{fig:componentswell_nosalt_6cycles}). At the end of the six cycles, Figure \ref{tab:lossAndprod_aege2023} shows that microbial activity has a weak impact on the production efficiency of the order of $-0.4\%$.\\
Furthermore, the total loss of \ce{H2}, the consumption of \ce{CO2} and the production of \ce{C1}, due to the microbial activity, are estimated comparing the total mass of \ce{H2} (resp. \ce{CO2}, \ce{C1}) over time in both cases, i.e. without and with microorganisms. At the end of the six cycles, microbial activity influences more or less the loss-production of gases: the loss of \ce{H2}  reaches $9.78\%$, the consumption of \ce{CO2} reaches $0.26\%$ while the production of \ce{C1} remains below $0.06\%$ (see Figure \ref{fig:components_well_lossproduct_6cycles} and table \ref{tab:lossAndprod_aege2023}).\\

It has to be noted that these simulations have been performed with two different linear solvers: LU factorization and ILU(0) preconditioned GMRES. As expected, the latter one is more efficient in terms of CPU time (see table \ref{tab:CPUTIME}).

\begin{table}[h!]
\centering
\scriptsize 
\caption{Test 3: loss and production (noted prod.) of gases due to the microbial activity in a pure and salt water aquifer and production efficiencies (noted Effic.).}
\label{tab:lossAndprod_aege2023}
\begin{tabular}{l l l l l l l l}
\toprule
Molecular Diffusion & $\msalt$ & $n_0$ &\ce{H2} loss  & \ce{CO2} loss& \ce{C1} prod.& Effic., no archae& Effic., archae\\
$(\unit{\meter\square\per\second})$&$(\unit{\mole\per\kgw})$ & $(\unit{\per\cubic\meter})$ &($\%$) &  ($\%$)& ($\%$)& ($\%$)& ($\%$)\\
\midrule
No&0&$10^8$ &$9.78 $& $0.26$  & $0.006$ & $94.28$ & $93.88$\\
No&0&$10^9$ &$47.63 $& $1.38$  & $0.12$ & $94.28$ & $90.13$\\
No&3&$10^8$ & $9.02$& $0.24$&$0.01$  & $94.27$ & $93.01$\\
No&3&$10^9$ & $46.63$& $1.35$&$0.12$  & $94.27$ & $90.42$\\
Yes&0&$10^8$ &$10.42 $& $0.3$  & $0.0033$ & $93.55$ & $93.15$\\
\bottomrule
\end{tabular}
\end{table}

\begin{table}[h!]
\centering
\scriptsize 
\caption{Test 3:  CPU times  using LU factorization and ILU(0) preconditioned GMRES.}
\label{tab:CPUTIME}
\begin{tabular}{l l l}
\toprule
Linear solver & Archae & CpuTime (\unit{\second}) \\
\midrule
 LU factorization&no archae &$973 $\\
 LU factorization&$10^8$ &$1263 $\\
 ILU(0) preconditioned GMRES&no archae &$320 $\\
ILU(0) preconditioned GMRES&$10^8$ &$471 $\\
\bottomrule
\end{tabular}
\end{table}

Secondly, we aim at studying the impact of the salinity on the microbial activity in UHS. For this, the previous
simulation cases are performed in a saline aquifer ($\msalt=3$). As expected, the salinity induces a decrease in the
solubility of \ce{H2} and \ce{CO2} independently of the presence of microorganisms. As a result the solubility of \ce{H2} (resp \ce{CO2}) in brine reaches $7.6\times 10^{-4}$ (resp $6.2\times 10^{-3}$) near the well, which is coherent with the results of section \ref{subsection:SW_EoS} in a salt water aquifer.\\
As a consequence, an increase in salinity produces a decrease in the solubility of \ce{H2} ($42\%$ in average) and \ce{CO2} ($36\%$ in average) which causes a decrease in the archae population ($1.8\%$ in average) as exhibited in Figure \ref{fig:pressure_nbact} (right).\\
Contrary to the microbial activity, the salinity neither influences the pressure of the reservoir nor the production efficiency. As a matter of fact, table \ref{tab:lossAndprod_aege2023} exhibits values in the pure water case  similar to those in the salt water case.\\
In the same way, the salinity shows very weak impact on the \ce{H2} loss (less than $1\%$), \ce{CO2} or $\ce{C1}$ production (Figure \ref{fig:components_well_lossproduct_6cycles} and table \ref{tab:lossAndprod_aege2023}).

Thirdly, we aim at studying the impact of the initial microbial density on the production efficiency and gases loss and production (see Figure \ref{fig:efficiencyH2_6cycles}). For this, we choose $n_0=10^{9}\unit{\per\cubic\meter}$. Figure \ref{fig:pressure_nbact} shows a significant decrease in the pressure and a microbial population following the rhythm of \ce{H2} injection-production. However an increase in the initial microbial density influences the long-term trend of the microbial population evolution, in so far as it is decreasing. It can be explained by the fact that archae population is too much numerous compared with the  \ce{H2} available in the liquid phase. Besides, with a lower initial archae density ($n_0=10^{8}\unit{\per\cubic\meter}$), the microbial population tends to increase until the end of the 6 cycles, but the slope becomes weaker along with time. So, we can guess that for more cycles, the average microbial population trend is liable to reach a threshold and then decline.\\
It has to be noted that with $n_0=10^{9}\unit{\per\cubic\meter}$, the loss of \ce{H2} and the consumption of \ce{CO2} reaches $11.01\%$ for the \ce{H2} and $0.45\%$ for the \ce{CO2}, at the end of the first cycle, which is consistent with those of \cite{Khoshnevis2023}. Moreover, the production efficiency reaches $61.76\%$  without organisms and  $60.98\%$ with a microbial activity, which are of the same order as the efficiencies obtained by  \cite{Khoshnevis2023} for 1 cycle. 
Finally, we aim at studying the effects of the molecular diffusion on the production efficiency and gases loss and production on a pure water aquifer. Independently of the microbial activity, the molecular diffusion tends to decrease the production efficiency $0.73\%$. In the same way, the molecular diffusion seems to impact weakly the loss-production of gases (see table \ref{tab:lossAndprod_aege2023}). 

\subsection{Test 4: 3D underground hydrogen storage with complex mixture}\label{Benchmark2022}

This test case is based on a large-scale 3D benchmark developed by the \href{https://www.ite.tu-clausthal.de/en/research/subsurface-energy-and-gas-storage/uhs-benchmark-study}{Institute of Subsurface Energy Systems (TU Clausthal)} and studied in detail by \cite{hogeweg2022benchmark}. The model represents a typical sandstone reservoir located at a depth of \(1210\unit{\meter}\), with horizontal dimensions of \(3050\unit{\meter} \times 3050\unit{\meter}\) and variable thickness. The domain is discretized using a corner-point grid consisting of \(61 \times 61 \times 12\) cells (a total of 44,652 cells). The reservoir has an average porosity of \(\Phi = 0.15\) and an absolute permeability tensor given by \(\mathbf{K} = [143\unit{\milli\darcy}, 143\unit{\milli\darcy}, 3\unit{\milli\darcy}]\).

A vertical injection well is placed at the center of the top surface, injecting a gas mixture rich in \ce{H2} (see Table~\ref{bench2022:gas_composition} for the full composition). 

\begin{table}[h]
    \centering
    \caption{Test 4: initial and injected gas compositions in mole fractions.}
    \begin{tabular}{lcc}
        \toprule
        Component & Initial (\%) & Injected (\%) \\
        \midrule
        Methane (\ce{CH4}) & 7.015 & 17.522 \\
        Water (\ce{H2O}) & 90.230 & 0.000 \\
        Ethane (\ce{C2H6}) & 0.270 & 0.144 \\
        n-Butane (\ce{C4H10}) & 0.040 & 0.000 \\
        Pseudo-component (\ce{C3+}) & 0.020 & 0.012 \\
        Hydrogen (\ce{H2}) & 0.000 & 80.000 \\
        Carbon Dioxide (\ce{CO2}) & 2.021 & 0.504 \\
        Nitrogen (\ce{N2}) & 0.405 & 1.818 \\
        \midrule
        Sum & 100.000 & 100.000 \\
        \bottomrule
    \end{tabular}
    \label{bench2022:gas_composition}
\end{table}

The system is initialized with a pressure of \(P_0 = 81.6\,\unit{\bar}\), a temperature of \(T_0 = 333.15\,\unit{\kelvin}\), and a uniform composition. Unlike the original setup, which uses depth-dependent initial compositions and phase distributions, our MRST implementation assumes spatially uniform initial conditions. All remaining model parameters are taken from \cite{hogeweg2022benchmark}.

The formulation of the overall composition is extended with microbial activity, resulting in approximately 401,000 degrees of freedom. To enhance computational performance, we employ the accelerated AD backend in MRST, together with the \texttt{AMGCL\_CPR}~\cite{demidov2019amgcl,NilsenAhmedCpr} preconditioner and a quasi-IMPES decoupling (block size = components + 1 for bacteria).

\begin{figure}
  \centering
  \includegraphics[width=0.482\linewidth]{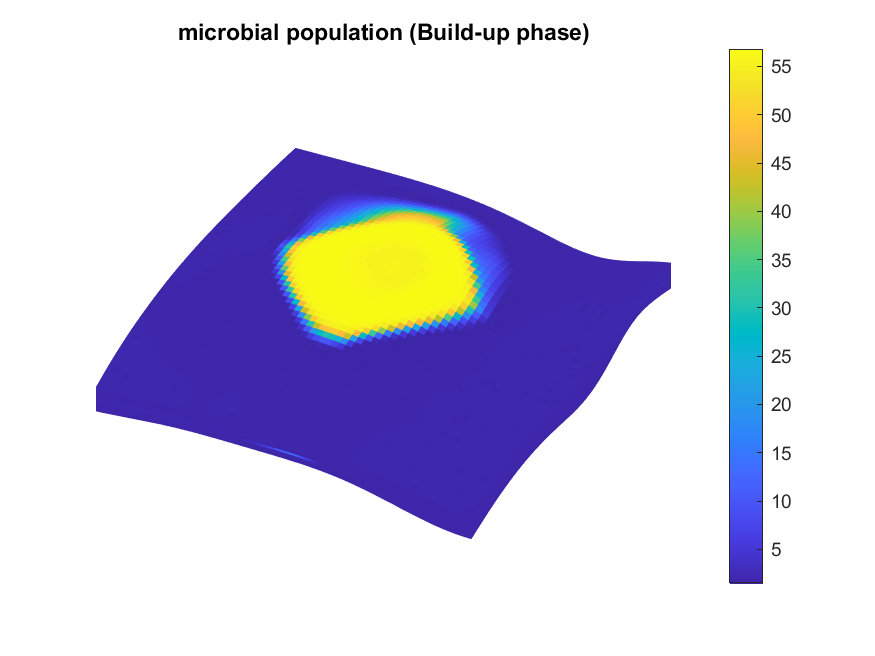}
  \includegraphics[width=0.4832\linewidth]{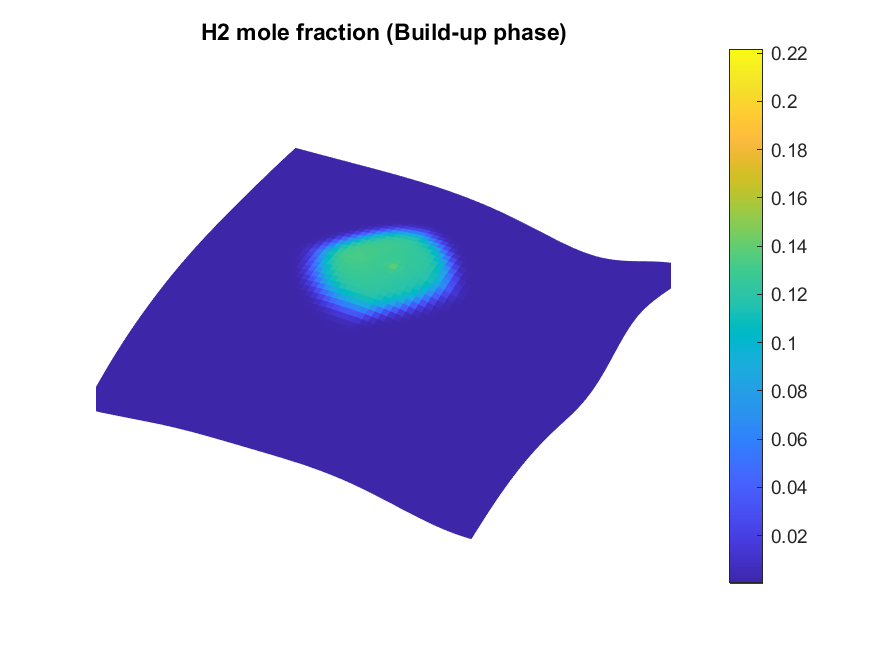}
  \caption{Test 4: Distribution of microbial population (left) and \ce{H2} mole fraction at the end of  build-up phase.}
  \label{fig_3d:microbial_and_gas_distribution}
\end{figure}

\begin{figure}
  \centering
  \includegraphics[width=0.4832\linewidth]{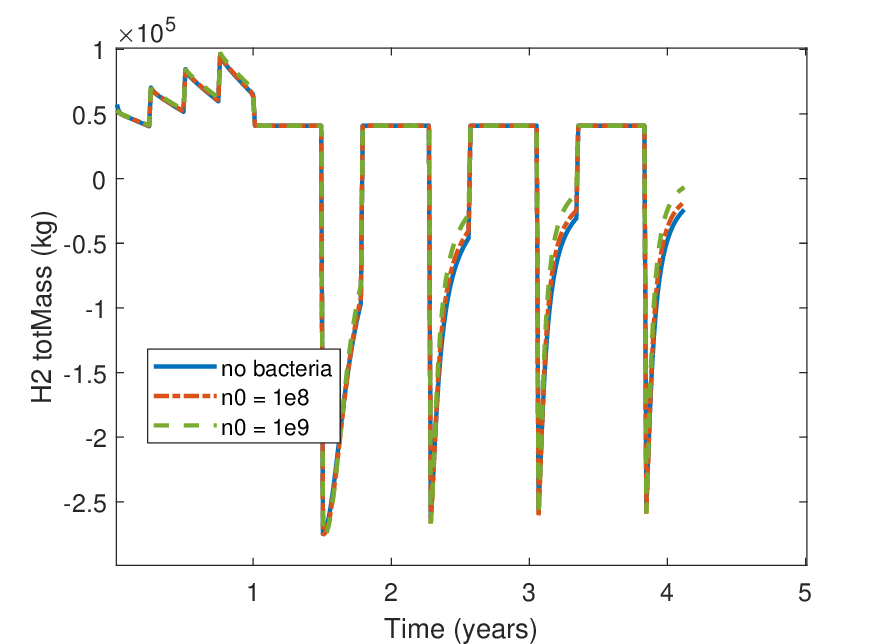}
  \includegraphics[width=0.4832\linewidth]{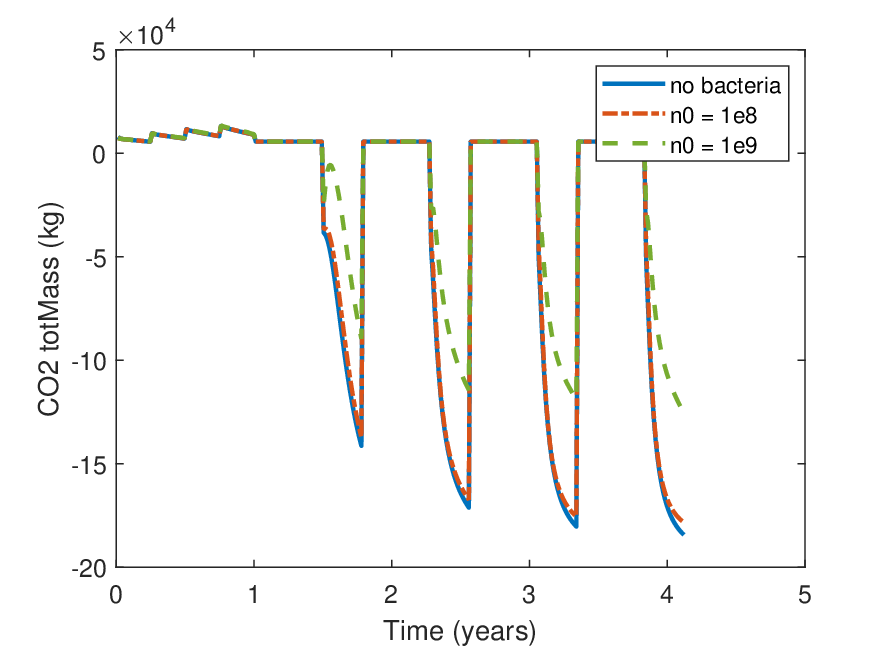}
  \includegraphics[width=0.4832\linewidth]{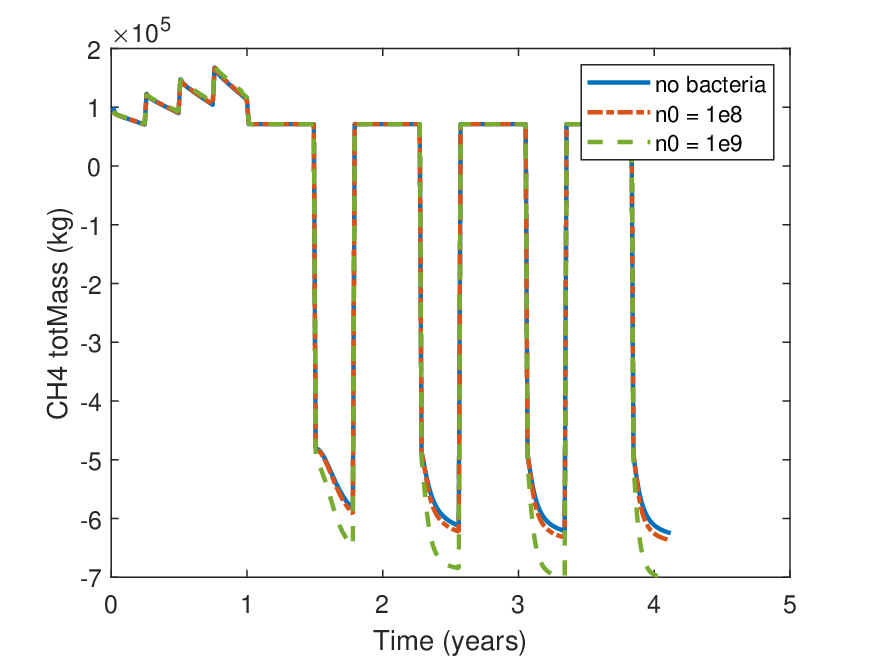}
  \includegraphics[width=0.4832\linewidth]{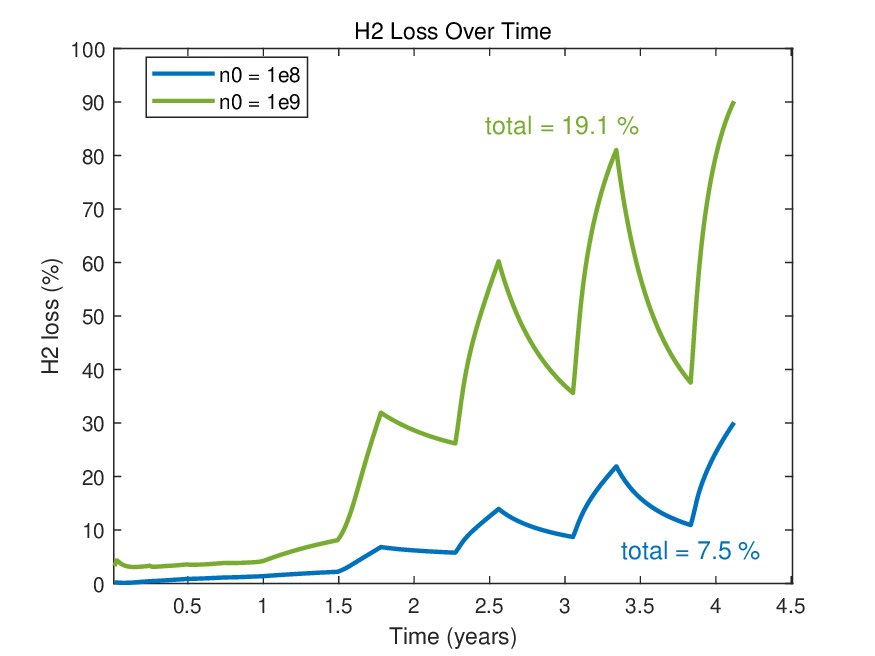}
  \caption{Test 4:  Impact of biochemical transformation on cyclic \ce{H2} injection. Total \ce{H2} loss amounts for 7.5 $\%$ for $n_0 =1.0e8$ and $19.1\%$ for $n_0 = 1.0e9$.}
  \label{fig_3d_bench:well_logs}
\end{figure}

This test focuses on the performance and behavior of simulations with and without microbial activity. For the latter, we consider two different initial microbial densities, \(n_0 \in \{10^8, 10^9\}\). Higher microbial concentrations intensify the coupling and increase the nonlinearity of the system. Figure~\ref{fig_3d:microbial_and_gas_distribution} depicts the spatial distribution of microbiel population (left) and \ce{H2} mole fraction (right) after the buil-up phase. As expected, both are concentrated near the well region where the \ce{H2} plume forms. Interestingly, the microbial plume extends slightly beyond the \ce{H2} plume, driven by the presence of even trace amounts of \ce{H2} and \ce{CO2} outside the core plume region. As illustrated in Figure~\ref{fig_3d_bench:well_logs} (bottom right), the total loss of \ce{H2} is approximately 7. 5\% for \(n_0 = 10^8\) and 19.1\% for \(n_0 = 10^9\). In the high-density case, the gas produced exhibits a significant impurity of methane (combined with a lower impurity due to \ce{CO2}), as seen in the bottom-left plot. In contrast, for \(n_0 = 10^8\), the gas remains relatively pure, although the trend suggests a gradual increase in impurity over time. In both cases, microbial activity did not affect \ce{H2} production, as shown in the top-left figure.
\begin{table}[h]
    \centering
    \caption{Test 4: comparison of simulation performance for abiotic and microbial cases.}
    \begin{tabular}{lccc}
        \toprule
        Metric & Abiotic & $n_0 = 10^8$ & $n_0 = 10^9$ \\
        \midrule
        Total Time [s]              & 1132  & 2382  & 5174 \\
        Time per Step [s]            & 3.7   & 7.8   & 16.9 \\
        Linear Its (Total)       & 574 &   3714  & 7366 \\
        Linear Its / step  (Mean) & 1.8   &  12.2  & 23.2 \\
        Nonlinear Its (Total)    & 494   & 1238  & 2135 \\
        Nonlinear Its / step (Mean)     & 1.9    & 4.2    & 7.4 \\
        \bottomrule
    \end{tabular}
    \label{tab:microbial_comparison}
\end{table}
Table~\ref{tab:microbial_comparison} reveals the important performance impact of including microbial activity. For \( n_0 = 10^8 \), the total simulation time increased from 1132 to 2382 (s) (\(+110\%\)), and further to 5174 (s) (\(+357\%\)) for \( n_0 = 10^9 \). This increase is attributed to the stronger nonlinear coupling introduced by microbial reactions, which led to \emph{more frequent time step cutting} and a significant rise in solver effort due to more demanding flash calculations, as the evolving composition required frequent and more complex updates at each step.

\section{Conclusion}

In this work, we have investigated the complex interplay of compositional, chemical, and biological processes involved in underground hydrogen storage (UHS). To this end, we have implemented in MRST: a SW EoS model to capture the solubility of gases, a double Monod model to describe microbial dynamics, a molecular diffusion model, and a bio-clogging model to assess the impact of microbial activity on rock properties.

We conducted numerous simulations involving liquid$–$gas phases and multicomponent systems with microbial activity. The simulation results are consistent with findings from the literature, particularly in showing the influence of salinity on the solubility of \ce{H2} and \ce{CO2}, which in turn affects microbial dynamics and hydrogen loss due to microbial consumption. This microbial activity subsequently alters rock properties, contributing further to \ce{H2} loss in the aquifer. Simulations show that hydrogen loss can reach up to \textbf{10\%} for initial microbial densities reported in the literature, increasing to \textbf{47\%} at higher densities. This is accompanied by \ce{H2} impurities from \ce{CH4} production. These results highlight the significant impact of microbial activity on hydrogen storage and underscore the need to incorporate bio-geochemical processes into UHS simulations.

Moreover, molecular diffusion effects were found to be relatively weak. However, since dispersive effects are more significant than diffusive ones, they should be incorporated in future work for a more comprehensive representation of transport phenomena.

\section*{Funding}
The authors affiliated with Sintef Digital acknowledge funding from the Norwegian Research and Innovation Centre for Hydrogen and Ammonia HYDROGENi (Grant No. 333118). Elyes Ahmed acknowledges support from \href{https://www.sintef.no/om-sintef/konsernsatsinger/}{KS Hydrogen: Konsernsatsinger i SINTEF for hydrogen}. The work of Brahim Amaziane and Stéphanie Delage Santacreu is partially supported by the ANR (French National Research Agency) within the HyStorEn project $n\textsuperscript{o}$ 22-CE05-0024, the Carnot Institute ISIFoR, and CNRS. The work of Stéphanie Delage Santacreu, Brahim Amaziane, Guillaume Galliéro and Salaheddine Chabab is carried out within the framework of the   \href{https://hydrogemm.cnrs.fr/}{GdR HydroGEMM :Hydrogène du sous-sol: étude intégrée de la Genèse... à la Modélisation Mathématique.} Their support is gratefully acknowledged.

\footnotesize
\bibliography{references}
\end{document}

%% file: comment-utils.tex
\usepackage{soul}

\soulregister\cite7
\soulregister\ref7
\soulregister\eqref7
\soulregister\num7
\soulregister\mysheader7
\soulregister\pageref7
\soulregister\myheader7
\soulregister\mycurrentheader7

\setul{0mm}{1mm}
\setstcolor{red}
\sethlcolor{yellow}

\ExplSyntaxOn
\cs_new:Npn \comprimitive:nn #1 #2 {\begingroup\protect\sethlcolor{#2}\hl{\myheader #1}\endgroup}
\NewDocumentCommand\createcomment {m m o}
{\IfValueTF {#3}
            {\str_set:Nx\l_tmpa_str {#3}
             \str_set:cx {#1header} {\str_use:N \l_tmpa_str}    
             \cs_new:cn {#1setheader:} {\def\myheader{\str_use:c{#1header} :~}}}
            {\str_set:Nn\l_tmpa_str {#1}
             \str_set:cx {#1header} {\str_uppercase:f {\str_head:N \l_tmpa_str} \str_tail:N \l_tmpa_str}
             \cs_new:cn {#1setheader:} {\def\myheader{\str_use:c{#1header} :~}}}
 \cs_new:cpn {#1textcom:n} ##1 { \use:c{#1setheader:} \comprimitive:nn { ##1} {#2}}
 \cs_new:cpn {#1mathcom:n} ##1 {\text{\use:c{#1textcom:n} {##1}}}
 \exp_args:Nc \NewDocumentCommand {#1com} {m} {\ifmmode \use:c{#1mathcom:n} {##1} \else \use:c{#1textcom:n} {##1}\fi  }
}
\ExplSyntaxOff

\colorlet{softgreen}{green!50!white}
\colorlet{softyellow}{yellow!50!white}
\colorlet{softred}{red!50!white}
\colorlet{softblue}{blue!50!white}